\documentclass[preprint,12pt]{elsarticle}

\usepackage[utf8]{inputenc}
\usepackage[T1]{fontenc}
\usepackage{amsmath,amsfonts,amssymb,booktabs}
\usepackage{graphicx}
\usepackage{subcaption}
\usepackage{textcomp}
\usepackage{url}
\usepackage{verbatim}
\usepackage{xcolor,soul,framed}
\usepackage{microtype}
\usepackage{float}
\usepackage[section]{placeins}
\usepackage{flafter}

\journal{Optics \& Laser Technology}

\begin{document}

\begin{frontmatter}

\title{Intensity Fluctuation Spectra as a Design Guide for Nonlinear-Tolerant Constellation Shaping}

\author[monash]{Ravneel Prasad}
\ead{ravneel.prasad@monash.edu}

\author[monash]{Emanuele Viterbo}
\ead{emanuele.viterbo@monash.edu}

\affiliation[monash]{organization={Department of Electrical and Computer Systems Engineering, Monash University},
            addressline={Clayton},
            postcode={VIC 3800},
            country={Australia}}

\begin{abstract}
Data-induced intensity fluctuations are an important driver of XPM-related nonlinear phase noise in coherent fiber links, particularly through their low-frequency spectral components. This paper develops a semi-analytical spectral framework that links block-level energy statistics of shaped constellations to the low-frequency features of the intensity-fluctuation power spectral density (PSD), thereby enabling spectral--temporal co-design for nonlinear mitigation. A semi-analytical PSD model is derived for finite-block-shaped symbols (including Constant Composition Distribution Matching (CCDM) and Enumerative Sphere Shaping (ESS)), explicitly exposing contributions from self-beating dependent on symbol energy variance, inter-symbol beating dependent on mean symbol energy, and block-induced energy variance terms. A compact expression for the spectral-dip width is obtained that captures the block length, symbol rate, pulse roll-off, and chromatic dispersion; this yields design rules for lowering the low-frequency content that most strongly drives the induced XPM. Resulting optimal symbol-rate laws are provided for shaped and unshaped systems and are validated by Monte Carlo simulations, which also confirm the distinct low-frequency behaviour of CCDM (suppressed DC) and ESS (finite DC pedestal at moderate block lengths). The framework consolidates prior time- and frequency-domain perspectives and provides actionable guidance for choosing block length, symbol rate, and shaping method to reduce nonlinear interference in high-capacity WDM systems.
\end{abstract}


\end{frontmatter}

\section{Introduction}

Constellation shaping has emerged as a key technique for improving the performance of optical communication systems \cite{skvortcov2020nonlinearity,askari2024probabilistic,amari2019introducing,geller2016shaping}. By tailoring the probability distribution or geometric positioning of transmitted symbols, shaping enables more efficient use of the available signal space, resulting in improved energy efficiency and reduced nonlinear impairments. In particular, shaping can provide up to 1.53 dB shaping gain in signal-to-noise ratio (SNR) for the additive white Gaussian noise (AWGN) channel \cite{amari2019introducing,geller2016shaping,dar2014shaping,guo2019multi}. However, with appropriate shaping, the aggregate shaping gain in a fiber-optic channel can exceed the ultimate gain achieved in an AWGN channel \cite{dar2014shaping,askari2024probabilistic}. This gain arises from a combination of reduced average energy, which improves the linear SNR, and modified energy statistics, which can reduce the low-frequency intensity fluctuations associated with nonlinear phase-noise penalties \cite{dar2014shaping,geller2016shaping,Prasad2025intenstiy}. 

Probabilistic shaping, in particular, has attracted considerable interest due to its compatibility with existing QAM systems without requiring changes to the constellation geometry \cite{vassilieva2022probabilistic}. It also offers rate adaptivity, which can be leveraged alongside fixed FEC to meet desired performance metrics across varying channel conditions \cite{vassilieva2022probabilistic,bocherer2015bandwidth}. At the heart of this process is the {\em distribution matcher} (DM), which maps information bits to a sequence of symbols adhering to a target distribution. While conventional Constant Composition Distribution Matching (CCDM) continues to advance through low-complexity multi-stage architectures for coherent transmission \cite{shi2025low}, practical implementations almost universally rely on fixed block lengths to constrain computational cost \cite{amari2019introducing,peng2021baud}. Consequently, numerous studies have investigated the impact of this finite block length on fiber nonlinearity \cite{amari2019introducing,geller2016shaping,skvortcov2020nonlinearity,fehenberger2020impact,askari2023probabilistic,fehenberger2020analysis}.

Cross-phase modulation (XPM) is strongly influenced by data-induced intensity fluctuations that evolve along a dispersive fiber \cite{du2011optimizing, lowery2022XPM}. The XPM efficiency itself exhibits low-pass characteristics, which depend on fiber length, dispersion, attenuation, and the wavelength separation between interacting channels. Consequently, the low-frequency components of these intensity fluctuations can become dominant contributors to XPM-induced phase noise \cite{lowery2022XPM,prasad2026linear}. Recently, in \cite{Prasad2025intenstiy}, intensity-fluctuation spectra were modeled for uniformly distributed symbols, and the composition of these spectra was analyzed as they evolved along the fiber. That study suggested that the intensity-fluctuation spectra could be altered by reducing beat components between neighboring symbols spaced $m$ symbols apart. A simple way of controlling the intensity spectra was demonstrated in \cite{chen2023penalty,zhang2023xpm}, where Manchester coding and direct-sequence spread-spectrum techniques were used to reduce low-frequency intensity fluctuations. This reduced the XPM induced by the coded system relative to an uncoded system and supported the view that low-frequency intensity fluctuations are important contributors to XPM-induced phase noise \cite{lowery2022XPM,prasad2026linear}.

A perspective from the standpoint of intensity fluctuations was provided in \cite{peng2021baud,peng2020transmission}. A super-symbol transmission scheme was proposed in \cite{peng2020transmission} that groups several 1D PAM components and maps them into 4D dual-polarization QAM super-symbols. Shaping in four dimensions, jointly over $(I_X,Q_X,I_Y,Q_Y)$, constrains the joint energy of each super-symbol and reduces instantaneous power variations across polarizations. This improves nonlinear tolerance compared to 1D shaping (independent amplitude shaping on $I$ and $Q$), 2D per-symbol QAM shaping, and systems without block-based shaping. Intuitively, moving from 1D to 2D to 4D increases the degrees of freedom available to control energy and thereby the temporal/frequency statistics of the power that drive fiber nonlinearities. The impacts of block length and symbol rate were later studied in \cite{peng2021baud}. These parameters were shown to create a spectral dip in the intensity fluctuation spectra near DC; this governed the nonlinear behavior, whereby the wider the spectral dip, the better the performance. 

In \cite{wu2021temporal}, the Energy Dispersion Index (EDI) was proposed to quantify the temporal unevenness of the signal energy. The approach partitions the waveform into fixed-duration windows, evaluates the energy within each window, and forms an index given by the average variance of these windowed energies normalized by their average energy level. EDI therefore measures ``burstiness", where smaller values indicate a more uniform energy distribution over time. As reported in \cite{wu2021temporal}, lower EDI values result in improved nonlinear tolerance in coherent fiber links, aligning with the intuition that reducing low-frequency power fluctuations mitigates nonlinear penalties. The EDI is a heuristic metric for quantifying the temporal unevenness of signal energy but does not provide any insight into the spectral characteristics of the intensity fluctuation spectra. The EDI did suggest that the statistics of the energy signal govern nonlinear performance and prompted \cite{dar2013properties} to show how fourth-order moments of data impact the phase noise. Following this, \cite{askari2024probabilistic} showed that the PSD of the energy signal was governed by the temporal statistical properties of the energy signal and explained their relationship to the phase noise via a perturbation-based linear filter model. The results presented in \cite{askari2024probabilistic} were similar to those presented in \cite{peng2021baud,peng2020transmission}, suggesting that low-frequency components are substantial contributors to nonlinear phase noise. 

Taken together, prior results point to two complementary but only partially connected views: (\textit{i}) time-domain statistics such as EDI and block-energy variability (set by shaping and DM block length), and (\textit{ii}) frequency-domain structure such as the DC notch/dip and its dependence on block length and symbol rate. What is still missing is a unified framework that \emph{maps concrete statistical descriptors of the energy signal} (mean of the per-block energy variance, variance of the mean energy across blocks, energy variance of the signal, and finite-length constraints) \emph{onto the spectral features} of the intensity-fluctuation PSD that most strongly drive XPM-induced phase noise, \emph{including how these mappings evolve under chromatic dispersion.} 

This work addresses this gap by developing a semi-analytical model that links the statistics of the energy signal to the spectral characteristics of the intensity-fluctuation PSD and to the induced phase variance. Specifically, we (\textit{i}) derive closed-form expressions that expose how finite block length, symbol rate, and DM type (e.g., Constant Composition Distribution Matching (CCDM) vs.\ Enumerative Sphere Shaping (ESS)) shape the low-frequency PSD features via terms such as the variance of mean block energy, energy variance across all symbols, and the mean of intra-block energy variance; (\textit{ii}) incorporate fiber dispersion to predict how the PSD evolves with distance; and (\textit{iii}) develop design rules that widen the low-frequency dip and minimize phase variance for given link lengths. The analysis is validated against MATLAB simulations that generate shaped QAM waveforms, apply pulse shaping, and propagate through a dispersion model. The resulting energy-signal PSDs and phase-variance trends agree with the semi-analytical predictions and with observations in \cite{peng2021baud,askari2024probabilistic}. Overall, the framework connects temporal and spectral viewpoints and provides practical guidance on selecting the block length and symbol rate, and on choosing between CCDM and ESS, to reduce low-frequency intensity fluctuations and reduce the impact of fiber nonlinearity.

\section{Intensity Fluctuations}
\subsection{Fundamentals on Intensity Fluctuations}
The optical intensity $I(t)$ is proportional to the squared magnitude of the optical electric field. For a linearly modulated signal, this intensity is represented as:
\begin{equation}\label{eq:Intensity}
    I(t) \propto \left| {A_{0}}{e^{j2\pi {f_{L}} t}} \cdot \left( \sum\limits_{k=-\infty}^{\infty} {{a_{{k}}} h(t-kT)} \right) \right|^{2}
\end{equation}
where $A_{0}$ is the peak amplitude of the optical carrier, $f_{L}$ is the laser frequency of the optical carrier, the $a_{k}$ are the data symbols taken from a zero-mean constellation, $h(t)$ is the pulse shaping function, and $T$ is the symbol period. In a dispersive fiber, the pulse broadens and, as a result, the intensity of the optical signal also evolves. To account for the dispersion-induced pulse broadening, the pulse shaping function is convolved with the fiber dispersion function
\begin{equation}\label{eq:pulsebroadening}
  s(t) = h(t) \otimes \mathcal{F}^{-1} (e^{j\beta_2 \omega^{2} z/2})
\end{equation} 
where $\beta_2$ is the fiber dispersion parameter, $\omega$ is the angular frequency, and $z$ is the distance along the fiber. 
Hence, \eqref{eq:Intensity} can be rewritten as:
\begin{equation}\label{eq:Intensity2}
  I(t) \propto \left| {A_{0}} {e^{j2\pi {f_{L}} t}} \cdot \left( \sum\limits_{k=-\infty}^{\infty} {{a_{{k}}} s(t-kT)} \right) \right|^{2}
\end{equation}

The intensity-fluctuation (IF) spectrum can be calculated by taking the Fourier transform of \eqref{eq:Intensity2}. Therefore, the IF spectrum can be expressed as:

\begin{equation}\label{eq:IntensityFluctuation}
  |I(f)|^2 \propto {\left| {{A_0}} \right|^4} {{B_a}(f)}
\end{equation}
where $B_a(f)$ is the power spectral density (PSD) of the energy signal of the pulse-shaped data sequence.

The PSD of the energy signal for a pulse-shaped sequence of independent and identically distributed (i.i.d.) zero-mean data symbols $a_k$ is given by \cite{Prasad2025intenstiy}:

\begin{subequations} \label{eq:SquaringSpectra}
\begin{align}
  B_{a}(f) &= \frac{\mu_{E}^{2}}{T^{2}} \sum_{l=-\infty}^{\infty} \left| \mathcal{G} \left( \frac{l}{T} ,0 \right) \right|^{2} \delta \left( f-\frac{l}{T} \right) \label{eq:FullModel_lines} \\
           &\quad + \frac{\sigma_{E}^{2}}{T} \left| \mathcal{G} (f,0) \right|^{2} \label{eq:FullModel_self} \\
           &\quad + \frac{2\mu_{E}^{2}}{T} \sum_{\substack{m=-\infty\\m\ne 0}}^{\infty} \left| \mathcal{G} (f,mT) \right|^{2} \label{eq:FullModel_neighbors}
\end{align}
\end{subequations}
where $E = |a|^2$ represents the symbol energy of a 1D symbol, $\mu_{E} = \mathbb{E}[E]$ is the mean energy of the data symbols, $\sigma_{E}^{2} = \mathbb{E}[E^{2}] - \mu_{E}^{2}$ is the variance of the energy of the data symbols, and $\mathcal{G}(f,\tau)$ is the Fourier transform of the pulse beat function $g(t,\tau)$ defined as:
\begin{equation}\label{eq:Gfunction}
  \mathcal{G}(f,\tau ) = \mathcal{F} [g(t,\tau )]
  \end{equation}
  \begin{equation}\label{eq:PulseBeat}
  g(t,\tau ) = s(t) \cdot {s^ * }(t - \tau )
\end{equation}
 
The above model, shown in \eqref{eq:SquaringSpectra}, reveals that the IF spectra comprise three components. The first, described by \eqref{eq:FullModel_lines}, consists of the spectral lines. The second component, given by \eqref{eq:FullModel_self}, describes the intensity spectrum resulting from each data pulse beating with itself. The third component, shown in \eqref{eq:FullModel_neighbors}, describes the spectra of the pulses beating with neighboring pulses that are spaced $mT$ apart.

In \cite{Prasad2025intenstiy}, the low-frequency IFs were explained to be due to the intensity spectrum of the subcarrier beating with itself in a multi-subcarrier system. Due to this, the IF spectrum due to a single subcarrier is used to explain the nonlinear performance of a shaped-constellation system.

\subsection{Energy-signal PSD for shaped data symbols}
The energy signal power spectral density (PSD) given in (\ref{eq:SquaringSpectra}) is valid under the assumption of i.i.d. zero-mean non-shaped data symbols. However, when distribution shaping is applied—particularly with finite block length constraints—the symbol energy exhibits structured fluctuations that alter its spectral characteristics. In such cases, the energy-signal PSD must be revised to account for correlations and energy variability introduced by the shaping process.

The modified energy-signal PSD for shaped data symbols with finite-length blocks and zero mean is given by:
\begin{eqnarray}\label{eq:1DPSD}
B_{a}(f) &=& \frac{\mu_{E}^{2}}{T^{2}} \sum_{l=-\infty}^{\infty} \left| \mathcal{G} \left( \frac{l}{T}, 0 \right) \right|^{2} \delta \left( f - \frac{l}{T} \right) \nonumber\\
&& + \frac{\sigma_{E}^{2}}{T} \left| \mathcal{G}(f, 0) \right|^{2} \nonumber\\
&& - \frac{\mu_{\sigma_{\mathrm{blk}}^{2}} - (n_s - 1)\sigma_{\mu_{\mathrm{blk}}}^{2}}{T} \left| \mathcal{H}(f, 0) \right|^{2} \nonumber\\
&& + \frac{2 \mu_{E}^{2}}{T} \sum_{\substack{m=-\infty\\m\ne 0}}^{\infty} \left| \mathcal{G}(f, mT) \right|^{2}
\end{eqnarray}
Here, $n_s$ is the number of symbols per block, and $K$ is the total number of blocks. The third term accounts for the energy variance structure imposed by shaping across and within blocks. Specifically,
$\mu_{\sigma_{\mathrm{blk}}^{2}}$ denotes the average of the intra-block symbol energy variances, and $\sigma_{\mu_{\mathrm{blk}}}^{2}$ denotes the variance of the inter-block mean energies. These are defined as:

\begin{eqnarray}
\mu_{\sigma_{\mathrm{blk}}^{2}} &=& \frac{1}{K} \sum_{k=1}^{K} \left( \frac{1}{n_s} \sum_{i=1}^{n_s} \left( E_i^{(k)} - \bar{E}^{(k)} \right)^2 \right) \\
\sigma_{\mu_{\mathrm{blk}}}^{2} &=& \frac{1}{K} \sum_{k=1}^{K} \left( \bar{E}^{(k)} - \bar{E} \right)^2
\end{eqnarray}

with
\begin{eqnarray}
\bar{E}^{(k)} = \frac{1}{n_s} \sum_{i=1}^{n_s} E_i^{(k)}, \quad
\bar{E} = \frac{1}{K} \sum_{k=1}^{K} \bar{E}^{(k)}
\end{eqnarray}

The finite-block correction is governed by the arithmetic mean, over all distinct ordered symbol-position pairs, of the covariance computed across the ensemble of shaping blocks.

\textbf{Lemma 1 (Block-Averaged Intra-Block Covariance of Symbol Energies).}
\textit{Let a block $k$ contain $n_s$ symbols with energies $E_i^{(k)}$. Let the variance of the block means be $\mathrm{Var}(\bar{E}^{(k)})=\sigma_{\mu_{\mathrm{blk}}}^{2}$, and let the average variance of the deviations within a block be $\mu_{\sigma_{\mathrm{blk}}^{2}}$. The arithmetic mean covariance over all distinct ordered symbol pairs $i\ne j$ within the same block is}
\begin{equation}\label{eq:lemma1_result}
\left\langle \mathrm{Cov}(E_i^{(k)},E_j^{(k)}) \right\rangle_{i\ne j}
=
\sigma_{\mu_{\mathrm{blk}}}^{2}
-
\frac{\mu_{\sigma_{\mathrm{blk}}^{2}}}{n_s-1},
\end{equation}
\textit{where}
\begin{equation}
\left\langle \mathrm{Cov}(X_i,X_j) \right\rangle_{i\ne j}
=
\frac{1}{n_s(n_s-1)}
\sum_{i=1}^{n_s}\sum_{\substack{j=1\\j\ne i}}^{n_s}
\mathrm{Cov}(X_i,X_j).
\end{equation}

\vspace{0.5em}
\noindent\textit{Proof.}
The energy of a symbol $E_i^{(k)}$ is decomposed into its block mean $\bar{E}^{(k)}$ and its deviation $\delta_i^{(k)}$:
\begin{equation}
E_i^{(k)}=\bar{E}^{(k)}+\delta_i^{(k)}.
\end{equation}
By construction, the deviations within each block satisfy
\begin{equation}
\sum_{i=1}^{n_s}\delta_i^{(k)}=0.
\end{equation}
Using the bilinear property of covariance and averaging over all distinct ordered pairs gives
\begin{align}
\left\langle \mathrm{Cov}(E_i^{(k)},E_j^{(k)}) \right\rangle_{i\ne j}
&=
\frac{1}{n_s(n_s-1)}
\sum_{i=1}^{n_s}\sum_{\substack{j=1\\j\ne i}}^{n_s}
\mathrm{Cov}
\left(
\bar{E}^{(k)}+\delta_i^{(k)},
\bar{E}^{(k)}+\delta_j^{(k)}
\right) \nonumber\\
&=
\sigma_{\mu_{\mathrm{blk}}}^{2}
+
\frac{1}{n_s(n_s-1)}
\sum_{i=1}^{n_s}\sum_{\substack{j=1\\j\ne i}}^{n_s}
\mathrm{Cov}\left(\bar{E}^{(k)},\delta_j^{(k)}\right) \nonumber\\
&\quad
+
\frac{1}{n_s(n_s-1)}
\sum_{i=1}^{n_s}\sum_{\substack{j=1\\j\ne i}}^{n_s}
\mathrm{Cov}\left(\delta_i^{(k)},\bar{E}^{(k)}\right) \nonumber\\
&\quad
+
\left\langle \mathrm{Cov}(\delta_i^{(k)},\delta_j^{(k)}) \right\rangle_{i\ne j}.
\end{align}

The first averaged cross-term is
\begin{align}
&\frac{1}{n_s(n_s-1)}
\sum_{i=1}^{n_s}\sum_{\substack{j=1\\j\ne i}}^{n_s}
\mathrm{Cov}\left(\bar{E}^{(k)},\delta_j^{(k)}\right) \nonumber\\
&\qquad =
\frac{1}{n_s(n_s-1)}
\sum_{j=1}^{n_s}
\sum_{\substack{i=1\\i\ne j}}^{n_s}
\mathrm{Cov}\left(\bar{E}^{(k)},\delta_j^{(k)}\right) \nonumber\\
&\qquad =
\frac{1}{n_s(n_s-1)}
\sum_{j=1}^{n_s}
(n_s-1)
\mathrm{Cov}\left(\bar{E}^{(k)},\delta_j^{(k)}\right) \nonumber\\
&\qquad =
\frac{1}{n_s}
\sum_{j=1}^{n_s}
\mathrm{Cov}\left(\bar{E}^{(k)},\delta_j^{(k)}\right) \nonumber\\
&\qquad =
\mathrm{Cov}
\left(
\bar{E}^{(k)},
\frac{1}{n_s}\sum_{j=1}^{n_s}\delta_j^{(k)}
\right) \nonumber\\
&\qquad =
\mathrm{Cov}\left(\bar{E}^{(k)},0\right)=0.
\end{align}
Similarly, the second averaged cross-term also vanishes by the same zero-sum constraint:
\begin{equation}
\frac{1}{n_s(n_s-1)}
\sum_{i=1}^{n_s}\sum_{\substack{j=1\\j\ne i}}^{n_s}
\mathrm{Cov}\left(\delta_i^{(k)},\bar{E}^{(k)}\right)
=
0.
\end{equation}
Therefore,
\begin{equation}\label{eq:cov_step1}
\left\langle \mathrm{Cov}(E_i^{(k)},E_j^{(k)}) \right\rangle_{i\ne j}
=
\sigma_{\mu_{\mathrm{blk}}}^{2}
+
\left\langle \mathrm{Cov}(\delta_i^{(k)},\delta_j^{(k)}) \right\rangle_{i\ne j}.
\end{equation}
To determine the averaged covariance between deviations, the zero-sum constraint is again applied:
\begin{equation}
\mathrm{Var}
\left(
\sum_{i=1}^{n_s}\delta_i^{(k)}
\right)=0.
\end{equation}
Expanding this variance gives
\begin{equation}
\sum_{i=1}^{n_s}\mathrm{Var}(\delta_i^{(k)})
+
\sum_{i=1}^{n_s}\sum_{\substack{j=1\\j\ne i}}^{n_s}
\mathrm{Cov}(\delta_i^{(k)},\delta_j^{(k)})
=0.
\end{equation}
The average deviation variance is defined as
\begin{equation}
\mu_{\sigma_{\mathrm{blk}}^{2}}
=
\frac{1}{n_s}
\sum_{i=1}^{n_s}
\mathrm{Var}(\delta_i^{(k)}).
\end{equation}
Hence,
\begin{equation}
n_s\mu_{\sigma_{\mathrm{blk}}^{2}}
+
n_s(n_s-1)
\left\langle \mathrm{Cov}(\delta_i^{(k)},\delta_j^{(k)}) \right\rangle_{i\ne j}
=0.
\end{equation}
Solving for the block-averaged covariance between deviations gives
\begin{equation}
\left\langle \mathrm{Cov}(\delta_i^{(k)},\delta_j^{(k)}) \right\rangle_{i\ne j}
=
-
\frac{\mu_{\sigma_{\mathrm{blk}}^{2}}}{n_s-1}.
\end{equation}
Finally, substituting this result into (\ref{eq:cov_step1}) yields
\begin{equation}
\left\langle \mathrm{Cov}(E_i^{(k)},E_j^{(k)}) \right\rangle_{i\ne j}
=
\sigma_{\mu_{\mathrm{blk}}}^{2}
-
\frac{\mu_{\sigma_{\mathrm{blk}}^{2}}}{n_s-1}.
\end{equation}

\vspace{0.5em}

\textbf{Lemma 2 (PSD Correction Coefficient).}
\textit{The per-symbol correction coefficient $C_{\mathrm{corr}}$, obtained by aggregating the block-averaged intra-block covariances over all distinct ordered pairs and normalizing by the block size $n_s$, is}
\begin{equation}\label{eq:lemma2_result}
C_{\mathrm{corr}}
=
(n_s-1)\sigma_{\mu_{\mathrm{blk}}}^{2}
-
\mu_{\sigma_{\mathrm{blk}}^{2}}.
\end{equation}

\vspace{0.5em}
\noindent\textit{Proof.}
The total covariance contribution over all $n_s(n_s-1)$ ordered pairs of distinct symbols within the block is calculated using the block-averaged covariance from Lemma~1:
\begin{align}
S_k
&=
n_s(n_s-1)
\left\langle \mathrm{Cov}(E_i^{(k)},E_j^{(k)}) \right\rangle_{i\ne j} \nonumber\\
&=
n_s(n_s-1)
\left[
\sigma_{\mu_{\mathrm{blk}}}^{2}
-
\frac{\mu_{\sigma_{\mathrm{blk}}^{2}}}{n_s-1}
\right] \nonumber\\
&=
n_s(n_s-1)\sigma_{\mu_{\mathrm{blk}}}^{2}
-
n_s\mu_{\sigma_{\mathrm{blk}}^{2}}.
\end{align}
To obtain the per-symbol contribution consistent with the standard PSD normalization, $S_k$ is normalized by the block size $n_s$:
\begin{equation}
C_{\mathrm{corr}}
=
\frac{S_k}{n_s}
=
(n_s-1)\sigma_{\mu_{\mathrm{blk}}}^{2}
-
\mu_{\sigma_{\mathrm{blk}}^{2}}.
\end{equation}

\vspace{1em}
\textbf{Remark 1 (Trivial Case $n_s=1$).} 
Consider the case where each block contains a single symbol ($n_s=1$). The intra-block deviation $\delta_1^{(k)}$ is zero, implying $\mu_{\sigma_{\mathrm{blk}}^2} = 0$. Substituting these values into Lemma 2 gives
\begin{equation}
C_{corr} = (1-1)\sigma_{\mu_{\mathrm{blk}}}^2 - 0 = 0.
\end{equation}
Thus, the block-correction term vanishes for this single-symbol block case. This limiting case is used only as a reference for the unshaped self-beating contribution; it is not intended to represent finite-block distribution matching or the general i.i.d. limit.

\vspace{1em}
The function $\mathcal{H}(f,\tau)$ in the third term of Equation (\ref{eq:1DPSD}) characterises the pulse interaction within each shaped block and is defined as the Fourier transform of $p(t,\tau)$:
\begin{gather}
\mathcal{H}(f, \tau) = \mathcal{F} \left[ p(t, \tau) \right]
\end{gather}
where
\begin{equation}
p(t, \tau) = u(t) \times u^*(t - \tau)
\end{equation}
Here, $u(t)$ denotes the dispersed block-level signal formed by linearly modulating a sequence of symbols, and the full block waveform is expressed directly as:

\begin{equation}\label{eq:blockenvelope}
u(t) = \frac{1}{\sqrt{n_{norm}}} \sqrt{\left(\sum_{i=1}^{n_s} \left|s(t - iT)\right|^2\right)}
\end{equation}

This scaling ensures that the block envelope has the same total energy as a single pulse, i.e.,
\begin{equation}
\int |u(t)|^2 dt = \int |s(t)|^2 dt
\end{equation}

This formulation ensures consistent energy scaling and allows accurate interpretation of block-induced fluctuations in the energy signal. In particular, the correction term involving $\mathcal{H}(f, 0)$ accounts for how block shaping suppresses low-frequency intensity variations by constraining the energy variations both within and across blocks.

\subsection{Energy-signal PSD for shaped data symbols - 2D symbols}
When considering 2D symbols, such as those used in quadrature amplitude modulation (QAM), the shaping process can introduce statistical dependence between the in-phase (I) and quadrature (Q) components. This dependence appears through the second-order moment $\mu_{IQ}=\mathbb{E}[IQ]$ and must be included in the energy-signal PSD analysis. Rather than computing separate PSD expressions for $I$ and $Q$, the 2D result can be written directly in terms of the symbol energy variance, the mean symbol energy, and the I/Q second moments. The power spectral density (PSD) of the energy signal for shaped 2D symbols can be expressed as:

\begin{eqnarray}\label{eq:2DPSD}
B_{a}(f) &=& \frac{\mu_{E}^{2}}{T^{2}} \sum_{l=-\infty}^{\infty} \left| \mathcal{G} \left( \frac{l}{T}, 0 \right) \right|^{2} \delta \left( f - \frac{l}{T} \right) \nonumber\\
&& + \frac{\sigma_{E}^{2}}{T} \left| \mathcal{G}(f, 0) \right|^{2} \\
&& - \frac{\mu_{\sigma_{\mathrm{blk}}^{2}} - (n_s - 1)\sigma_{\mu_{\mathrm{blk}}}^{2}}{T} \left| \mathcal{H}(f, 0) \right|^{2} \nonumber\\
&& + \frac{ \mu_{E}^{2}+ (\mu_{I^2} -\mu_{Q^2})^{2} + 4{\mu^2_{IQ}}}{T} \sum_{\substack{m=-\infty\\m\ne 0}}^{\infty} \left| \mathcal{G}(f, mT) \right|^{2} \nonumber
\end{eqnarray}

The coefficient of the fourth term follows from the complex second moment of $a=I+jQ$. For independent symbols drawn from the same 2D constellation, the nonzero inter-symbol beating contribution contains
\begin{equation}
\mathbb{E}[|a|^2]^2+\left|\mathbb{E}[a^2]\right|^2.
\end{equation}
Since
\begin{equation}
\mathbb{E}[a^2]=\mathbb{E}\{(I+jQ)^2\}=\mu_{I^2}-\mu_{Q^2}+j2\mu_{IQ},
\end{equation}
we obtain
\begin{equation}
\left|\mathbb{E}[a^2]\right|^2=(\mu_{I^2}-\mu_{Q^2})^2+4\mu_{IQ}^2,
\end{equation}
which gives the fourth-term coefficient $\mu_E^2+(\mu_{I^2}-\mu_{Q^2})^2+4\mu_{IQ}^2$.

In this expression, $E = |a|^2 = I^2 + Q^2$ represents the energy of a complex 2D symbol $a$, where $\mu_{E} = \mathbb{E}[E]$ is the mean symbol energy and $\sigma_{E}^{2}$ is its variance. The terms $\mu_{I^2} = \mathbb{E}[I^2]$ and $\mu_{Q^2} = \mathbb{E}[Q^2]$ denote the average power of the in-phase ($I$) and quadrature ($Q$) components, respectively, while other parameters retain their definitions from the 1D analysis. The fourth term of the equation differs from the 1D case in Equation~(\ref{eq:1DPSD}). Its new form, derived from the properties of complex random variables, introduces the expression $(\mu_{I^2} - \mu_{Q^2})^2 + 4\mu_{IQ}^2$, which explicitly accounts for any power imbalance and correlation between the $I$ and $Q$ components. For unshaped data, the third term, which is associated with the block structure, vanishes. 

\section{Simulation Setup}
The analytical models for the energy-signal PSD in Equations~(\ref{eq:1DPSD}) and (\ref{eq:2DPSD}) are first validated against numerical simulations in MATLAB. Two shaping implementations are considered: constant-composition distribution matching (CCDM), which directly imposes a fixed target symbol distribution, and enumerative sphere shaping (ESS), which indirectly selects sequences subject to a maximum block-energy constraint. ESS typically achieves lower rate loss than CCDM for a given shaping rate \cite{amari2019introducing,gultekin2019enumerative}, although it does not directly control symbol-distribution entropy. For validation, shaped QAM symbols are generated, filtered by a root-raised-cosine (RRC) filter with roll-off 0.1, and their PSD is computed by Fourier transform. The framework is then extended with a fiber-dispersion model to verify the analytical predictions in the presence of chromatic dispersion.

System-level validation is performed in VPItransmissionMaker Optical Systems v11.5. The transmitter generates shaped QAM symbols, pulse-shaped by an RRC filter with roll-off 0.05, and linearly modulated onto an optical carrier with launch power $1$ mW. The signal propagates through an 80-km Standard Single-Mode Fiber (SSMF) span with dispersion 16 ps/nm/km, nonlinear index $2.6\times10^{-20}$ m$^2$/W, and effective area $80~\mu\text{m}^2$. This setup validates the intensity-fluctuation model in Equation~(\ref{eq:IntensityFluctuation}) and assesses the effect of shaping-induced spectral changes on XPM. XPM phase fluctuations are isolated using a two-channel configuration with 50 GHz spacing and a baud rate of 32~GBd, consisting of a shaped QAM pump channel ($1$ mW) and a continuous-wave probe channel ($10~\mu$W) used to capture the phase fluctuations.

This work focuses on isolating the relation between finite-block shaping, low-frequency intensity-fluctuation spectra, and pump-induced XPM phase fluctuations. The optical validation therefore uses a shaped pump channel and a low-power continuous-wave probe to isolate XPM-induced phase variations. The model is not intended to replace full GN/EGN or NLSE-based system models including ASE noise, polarization multiplexing, carrier recovery, or full WDM coded-modulation performance. Rather, it provides a semi-analytical design tool for understanding how block-level energy statistics influence the low-frequency spectral components that drive XPM.

\section{Numerical Results}
\subsection{Energy-signal PSD of Shaped Symbols}
\subsubsection{Impact of Shaping Methods}
In analyzing the energy–signal PSD for shaped QAM symbols and its impact on nonlinear tolerance, two complementary probabilistic‑shaping approaches are commonly used \cite{gultekin2019enumerative,amari2019introducing,askari2024probabilistic}: \emph{direct shaping} and \emph{indirect shaping}. Direct shaping imposes the target symbol probabilities explicitly. A practical realization is constant‑composition distribution matching (CCDM), which maps information bits to fixed‑length blocks so that the empirical histogram closely follows a prescribed probability mass function within the limits of finite blocklength. For a QAM alphabet and a block of length $n_s$, a composition specifies how many times each constellation point appears; all sequences with that composition are treated as equally likely. This enforces the desired marginal distribution and reduces inter‑block variability in symbol counts. On the other hand, indirect shaping does not set symbol probabilities directly. Instead, it selects sequences according to a block‑energy constraint: bits are mapped uniformly among all length-$n_s$ sequences whose total energy (sum of squared symbol magnitudes over the block) lies within a prescribed bound, with a shell variant using an equality constraint. The resulting marginal distribution emerges from the uniform measure on this constrained set and typically approximates an energy‑efficient profile. 

\begin{figure}[!htbp]
  \centering
  \begin{subfigure}[t]{0.76\linewidth}
    \centering
    \includegraphics[width=\linewidth]{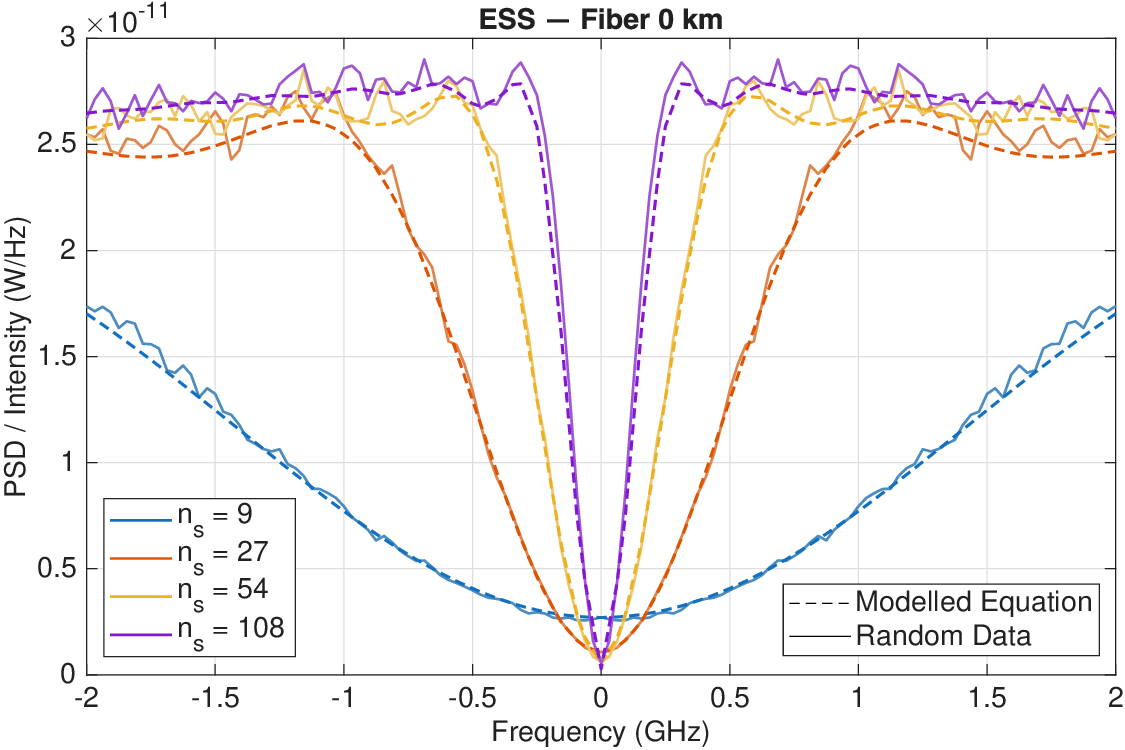}
  \end{subfigure}
  \hfill
  \begin{subfigure}[t]{0.76\linewidth}
    \centering
    \includegraphics[width=\linewidth]{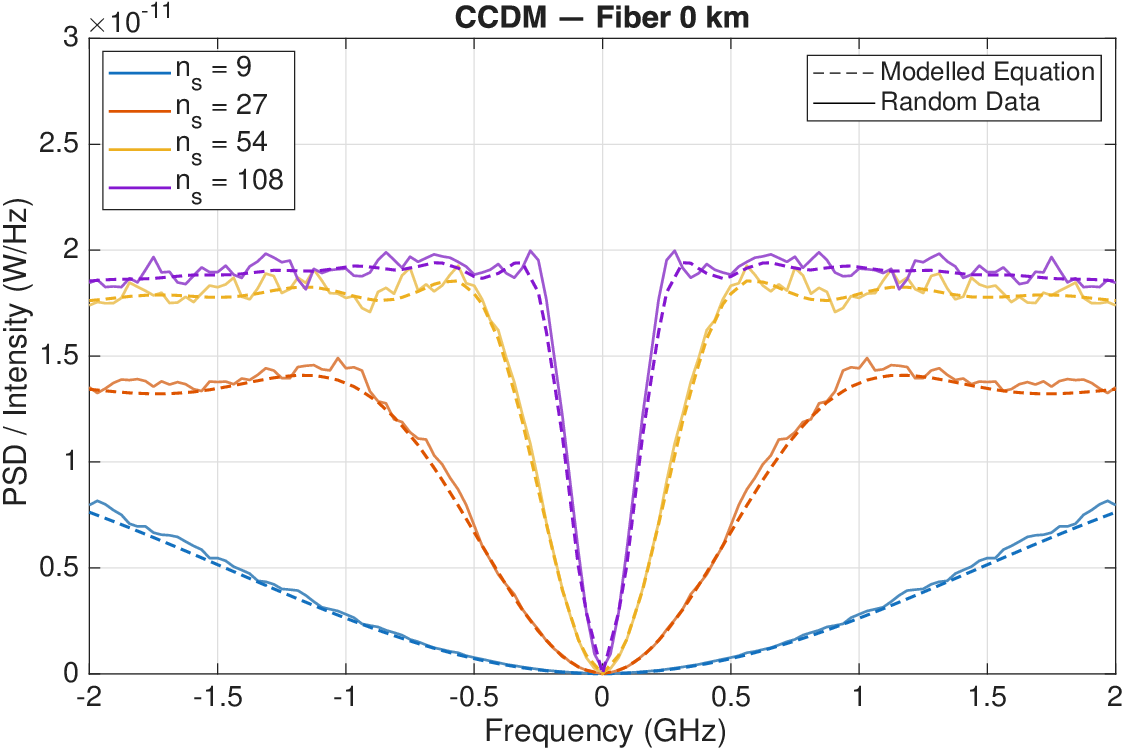}
  \end{subfigure}
  \caption{Energy-signal PSDs for shaped QAM symbols: ESS (top) and CCDM (bottom) for different block lengths. The shaped constellations are normalized to have the same mean energy.}
  \label{fig:psd-comparison}
\end{figure}

To verify the energy-signal PSD for shaped QAM symbols, two shaping algorithms are considered: CCDM with a Maxwell-Boltzmann target distribution to generate the desired symbol probabilities (or block composition), and Enumerative Sphere Shaping (ESS), for which shell mapping is applied pragmatically. Their PSD behaviours differ, especially at the low frequencies discussed in \cite{amari2019introducing,askari2024probabilistic}. This can be seen in Fig.~\ref{fig:psd-comparison}, where the PSDs of the energy signals for both CCDM and ESS are plotted for different block lengths before dispersion. The PSDs are normalized to have the same mean energy, allowing a direct comparison of their spectral characteristics. The ESS PSD exhibits more pronounced energy fluctuations at $f = 0$~Hz for shorter block lengths, whereas the CCDM PSD shows more consistent behaviour across different block lengths, indicating that the energy fluctuations are zero at $f = 0$~Hz. The exact DC component corresponds to a constant phase offset after XPM and is therefore not itself a time-varying phase-noise component. However, the finite low-frequency pedestal around DC remains important because the XPM transfer function is strongest at low frequencies. Thus, the DC value reported in the PSD is used here as an indicator of the near-DC pedestal that drives the phase-variance trends, not as a fluctuating phase term by itself. The PSDs are generated for a 256-QAM constellation pulse-shaped with a root-raised-cosine (RRC) filter with a roll-off factor of 0.1. 

The same interpretation also clarifies how advanced ESS variants that explicitly reduce block-energy variations, the fourth-order moment, or kurtosis would affect the spectra. Here, the fourth-order moment/kurtosis should not be viewed as an independent parameter separate from the symbol-energy variance. Since $E=|a|^2$, the symbol-energy variance is $\sigma_E^2=\mathbb{E}[E^2]-\mu_E^2=\mathbb{E}[|a|^4]-\mu_E^2=\mu_E^2(\kappa-1)$, where $\kappa=\mathbb{E}[|a|^4]/\mu_E^2$ is the normalized fourth-order moment. Thus, for shaped constellations normalized to the same mean energy, reducing the fourth-order moment or kurtosis directly changes $\sigma_E^2$. Such variants can be incorporated in the present model without changing the PSD derivation: their effect enters through the recomputed values of $\sigma_E^2$, $\mu_{\sigma_{\mathrm{blk}}^{2}}$, $\sigma_{\mu_{\mathrm{blk}}}^{2}$, and the I/Q second moments. If an ESS variant reduces inter-block mean-energy fluctuations, the finite near-DC pedestal is expected to decrease and its behaviour would move closer to that of CCDM. If it reduces the constellation fourth-order moment, the symbol-energy-variance and within-block-variation terms are correspondingly reduced. Therefore, the proposed framework provides a direct way to compare standard ESS and energy-variation-reduced ESS designs through their measured or predicted block-energy statistics.

\subsubsection{Spectral Composition of the Energy-signal PSD}

Equation~(\ref{eq:2DPSD}) provides a theoretical framework for calculating the energy-signal PSD of shaped QAM symbols. This model accurately reflects the spectral characteristics observed in Fig.~\ref{fig:psd-comparison}, including the reduction of low-frequency energy fluctuations due to shaping. The agreement between the analytical model and the simulated PSDs confirms the validity of the approach for describing the impact of probabilistic shaping on the energy signal spectrum. Since the PSD is well described by Equation~(\ref{eq:2DPSD}), it can be used to analyze the spectral composition of the energy signal for shaped QAM symbols. Based on Equation~(\ref{eq:2DPSD}), the spectrum is composed of four different components: the first term represents the spectral lines of the energy signal, the second term represents the energy of the data symbols beating with themselves, the third term represents the component resulting from the shaping (or introduction of block-based shaping), and the fourth term represents the energy of the data symbols beating with themselves. Here, the spectral lines are omitted since the main focus is on understanding the energy fluctuations at low frequencies. Signal shaping introduces block-based energy variance, which is reflected in the third term of Equation~(\ref{eq:2DPSD}). This term accounts for the energy variance structure imposed by shaping across and within blocks, which is crucial for understanding how shaping affects the spectral characteristics of the energy signal. This reduces the energy fluctuations produced by the self-beating of the symbols, which is the major contributor to the low-frequency components below 2 GHz. Although the spectral composition shown is for an ESS-shaped signal, it is valid for other shaping methods as long as the statistical properties outlined in Equation~(\ref{eq:2DPSD}) are known. For CCDM-shaped symbols, the variance of the mean energy between blocks is zero ($\sigma_{\mu_{\mathrm{blk}}}^{2} = 0$), while the mean energy variance within blocks is equivalent to the collective energy variance of the symbols ($\mu_{\sigma_{\mathrm{blk}}^{2}} = \sigma_{E}^{2}$). As a result, the energy fluctuations at $f = 0$ Hz are totally suppressed, which is evident in Fig.~\ref{fig:psd-comparison}. However, for uniformly distributed symbols, the effect of the third term in Equation~(\ref{eq:2DPSD}) is negligible, and sub-2 GHz fluctuations are therefore always present unless the symbol energy variance is zero, which occurs only for a QPSK modulation format. This indicates that the shaping process significantly alters the spectral characteristics of the energy signal, particularly at low frequencies.

\begin{figure}[!htbp]
  \centering
  \includegraphics[width=0.76\linewidth]{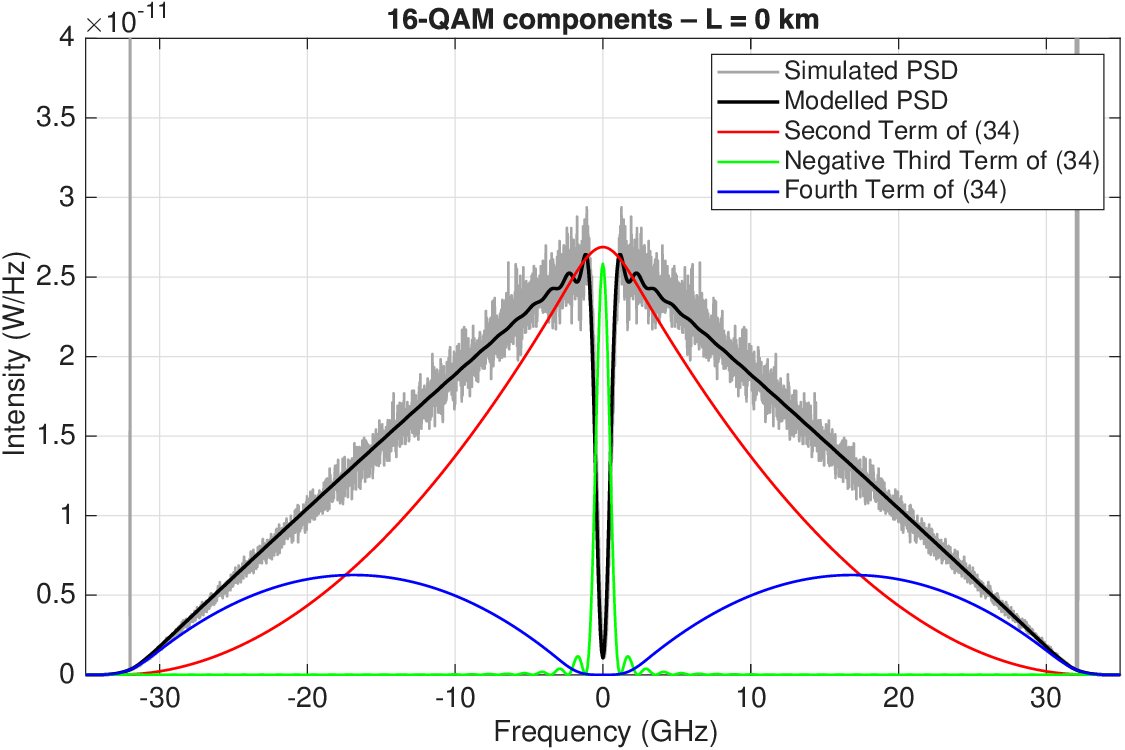}
  \caption{Spectral composition of the energy-signal PSD for ESS-shaped QAM symbols with $n_s=27$, based on Equation~(\ref{eq:2DPSD}).}
  \label{fig:PSDComposition}
\end{figure}

\subsubsection{Impact of Dispersion}
\begin{figure}[!htbp]
  \centering
  \begin{subfigure}[t]{0.76\linewidth}
    \centering
    \includegraphics[width=\linewidth]{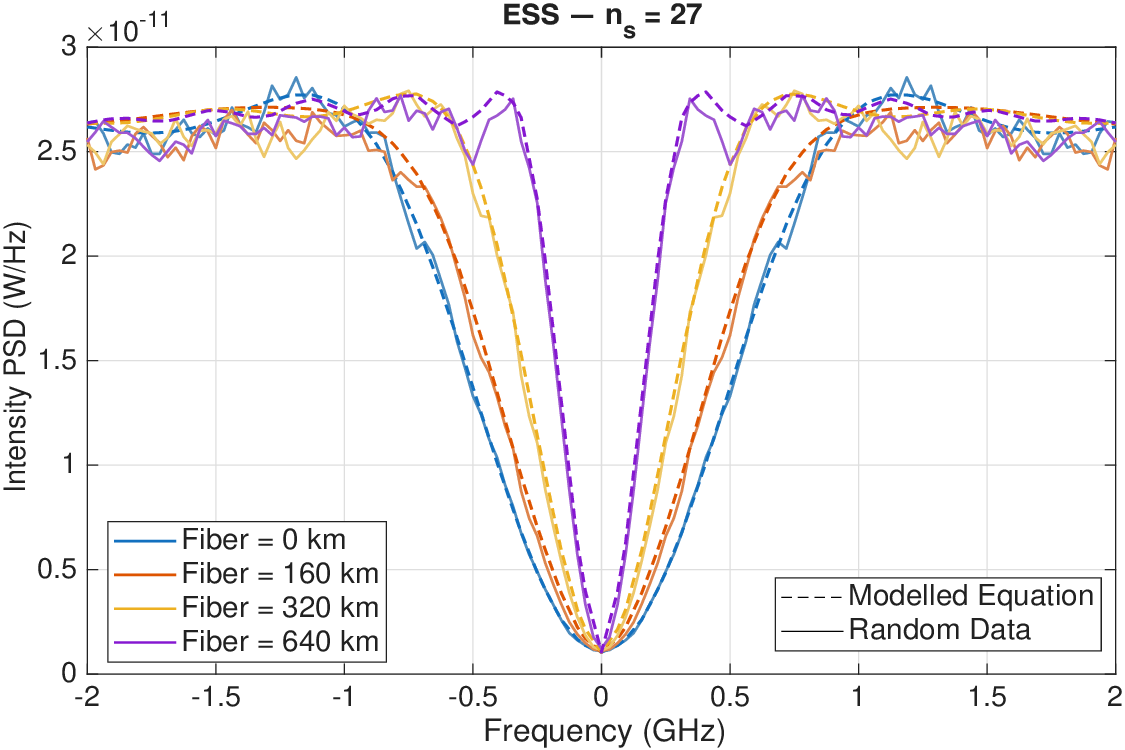}
  \end{subfigure}
  \hfill
  \begin{subfigure}[t]{0.76\linewidth}
    \centering
    \includegraphics[width=\linewidth]{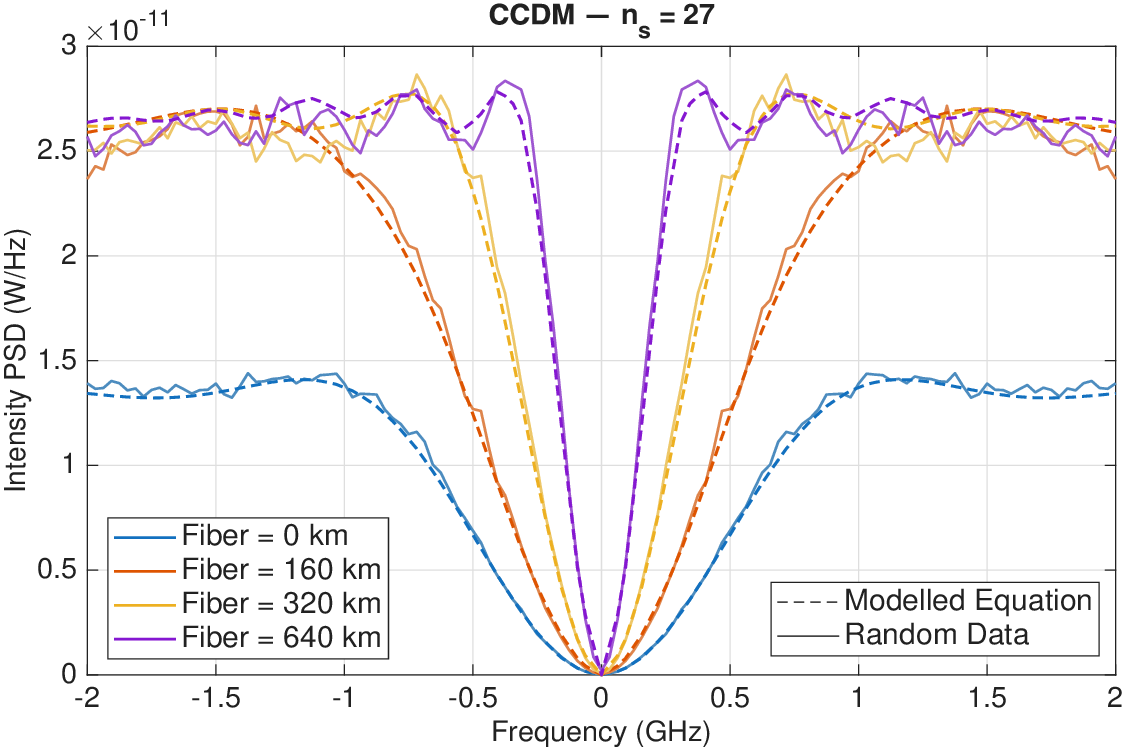}
  \end{subfigure}
  \caption{Energy-signal PSDs for shaped QAM symbols: ESS (top) and CCDM (bottom) for different fiber lengths in the dispersion model. The block length is set to 27 symbols per block.}
  \label{fig:psddistance-comparison}
\end{figure}

The energy-signal PSD is proportional to the intensity-fluctuation spectrum, as indicated in Equation~(\ref{eq:IntensityFluctuation}). Therefore, when the pulse-shaped signals of the shaped QAM symbols pass through a dispersion model, the resulting change in the energy-signal PSD is captured by Equation~(\ref{eq:2DPSD}), as shown in Fig.~\ref{fig:psddistance-comparison}. The results illustrate that the intensity fluctuations tend to grow with transmission length in the dispersion model described by Equation (\ref{eq:pulsebroadening}). Block length appears to have an effect similar to that of symbol rate: the PSD tends to increase with transmission length, as established in \cite{peng2021baud}. It is important to note that these are not intensity-fluctuation spectra, since the energy-signal PSD of the shaped QAM symbols needs to be scaled appropriately to obtain the intensity-fluctuation spectra described by Equation (\ref{eq:IntensityFluctuation}). Therefore, only the effect of the dispersion function on the pulse-shaped signals of the shaped QAM symbols is considered here.

\begin{figure}[!htbp]
  \centering
  \includegraphics[width=0.76\linewidth]{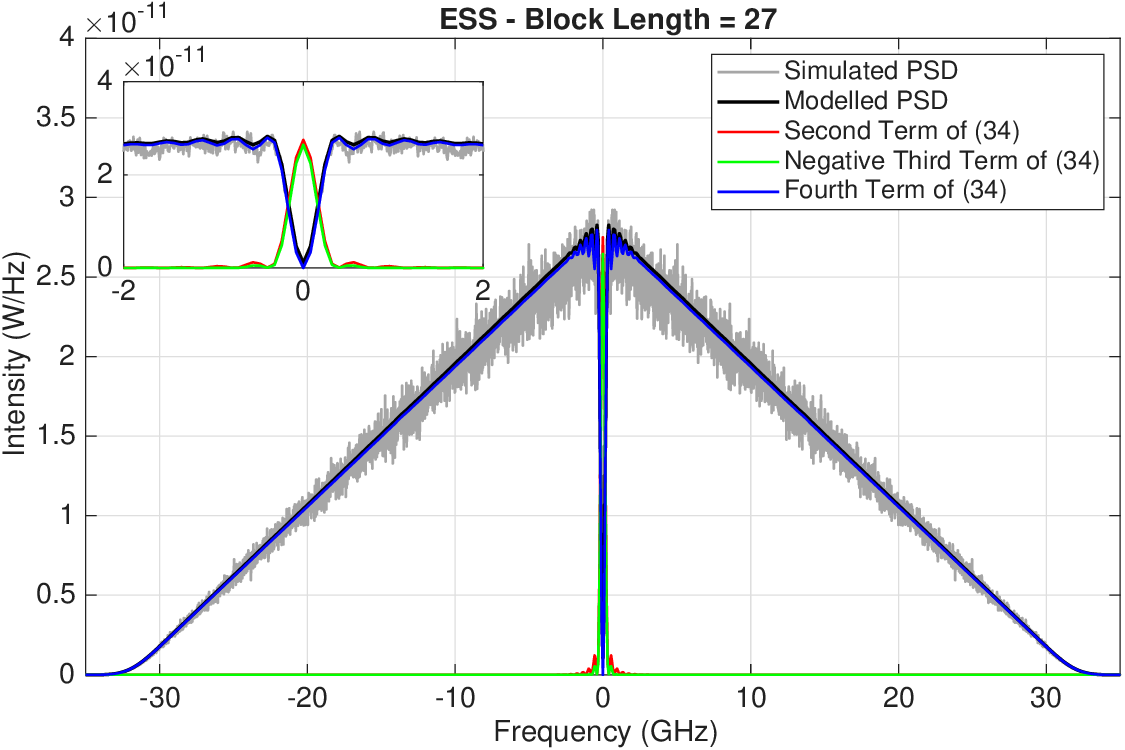}
  \caption{Spectral composition of the energy-signal PSD for ESS-shaped QAM symbols with $n_s=27$ and a fiber length of 640~km, based on Equation~(\ref{eq:2DPSD}).}
  \label{fig:PSD640KComposition}
\end{figure}

The increase in the energy-signal PSD is mainly due to the increased pulse overlap caused by accumulated dispersion along the fiber. This is evident in the spectral composition of the energy-signal PSD for ESS-shaped QAM symbols with the block length set to 27 in Fig.~\ref{fig:PSD640KComposition}. The spectral component due to the fourth term in Equation~(\ref{eq:2DPSD}) becomes a major contributor to the energy fluctuations and has grown significantly relative to the corresponding component in Fig.~\ref{fig:PSDComposition}, which has not undergone dispersion. The spectral components due to the second and third terms in Equation~(\ref{eq:2DPSD}) are still present, but their bandwidths are significantly reduced, which is reasonable because pulse broadening reduces the bandwidth of the energy signal spectrum. However, the peak at $f = 0$ Hz does not change. This is in line with the findings in \cite{Prasad2025intenstiy}.

\subsection{Spectral Dip Width}
\subsubsection{Analytical Formulation}
Now that it has been established that the spectral dip is entirely due to the third term of Equations~(\ref{eq:1DPSD}) and (\ref{eq:2DPSD}), the width of the spectral dip can be controlled by adjusting the block length and the symbol rate of the signal. Considering this, the duration of this block envelope before dispersion can be represented as:

\begin{equation}\label{eq:BlockDuration}
  T_{b} = (n_{s}-1 + 2\,\frac{t_z}{T})\,T  = \frac{n_{s} - 1 + 2\,a}{R_{s}}
\end{equation}
where $t_z$ is the main-lobe half-width of the shaped pulse, $a$ is the ratio of this half-width to the symbol period (i.e., $a = {t_z}{R_s}$), and $R_{s}$ is the symbol rate for the shaped QAM symbols. The value of $a$ is determined by the pulse shape: for an RC pulse $a=1$, for a rectangular pulse $a=0.5$, and for an RRC pulse $a$ depends on $\beta$ but is approximately~1 for small $\beta$ values. 

The width of the spectral dip at $f=0$~Hz is inversely proportional to the block length $n_s$, as indicated by the third term in Equations~(\ref{eq:1DPSD}) and (\ref{eq:2DPSD}). This dip is crucial, as it suppresses the low-frequency intensity fluctuations (IFs) that contribute to XPM phase noise.

Chromatic dispersion introduces a similar effect. As the signal propagates, the block envelope is stretched, which narrows the spectral dip by reducing the bandwidth of the envelope's energy PSD. This dispersion-induced narrowing, evident in Fig.~\ref{fig:psddistance-comparison} as the fiber length increases, reduces the frequency range over which low-frequency IFs are suppressed. The duration of the block envelope after dispersion is given by:
\begin{equation}
  T_{b}^{\prime} = \frac{n_{s} -1 + 2\,a}{R_{s}} + DLR_{s}\lambda^{2} /c
\end{equation}
Here, $D$ is the dispersion coefficient, $L$ is the fiber length, $\lambda$ is the wavelength of the signal, and $c$ is the speed of light in vacuum. 

To account for pulse shaping with roll-off $\,\beta\,$ (e.g., RRC), a multiplicative factor $\kappa_\beta$ is introduced to capture the bandwidth expansion of the baseband signal. For the block envelope, $\kappa_\beta \approx 1+\beta$ is used ($\kappa_\beta=1$ recovers the rectangular-pulse limit). After dispersion, the block-envelope duration becomes
\begin{equation}\label{eq:BlockDurationAfterDispersion}
  T_b' \;=\; \frac{n_s - 1 + 2\,a}{R_s} \;+\; \kappa_\beta\, D\,L\,R_s\,\frac{\lambda^2}{c},
\end{equation}

The bandwidth of the envelope is given by the inverse of Equation~(\ref{eq:BlockDurationAfterDispersion}):
\begin{equation}\label{eq:SpectralDipWidth}
  \Delta f_{b} = \frac{2}{T_{b}^{\prime}} 
\end{equation}

\begin{figure}[!htbp]
  \centering
  \includegraphics[width=0.76\linewidth]{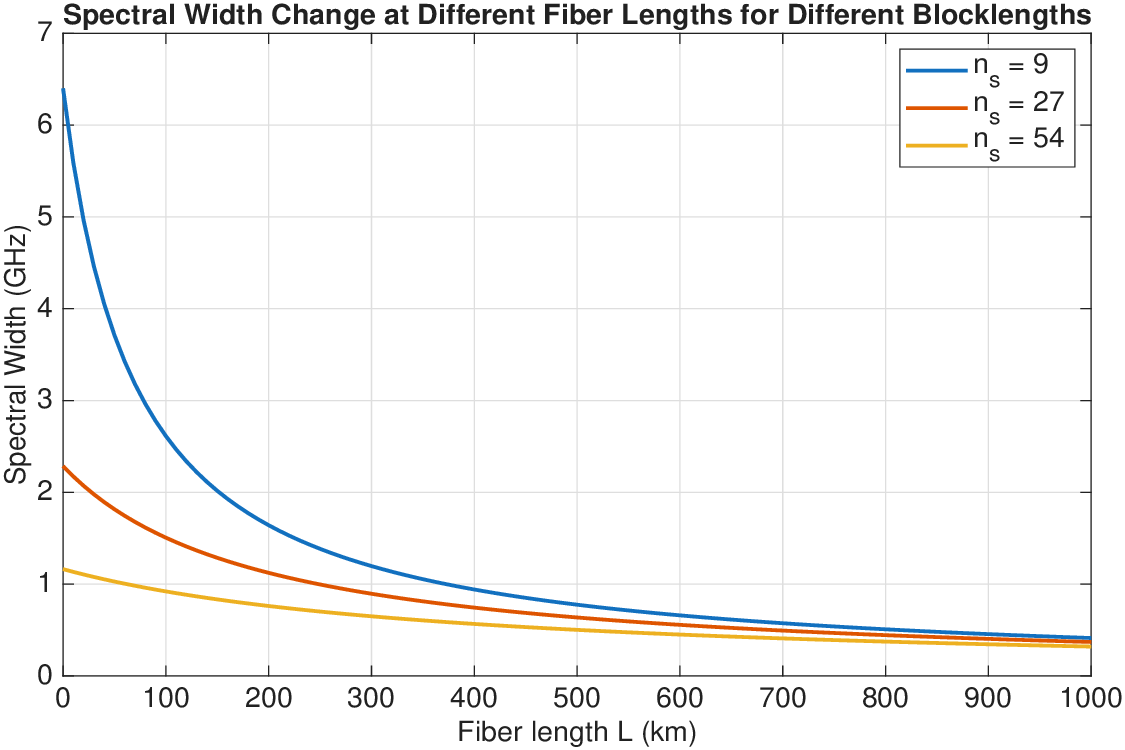}
  \caption{Modeled width of the spectral dip at $f=0$~Hz for different block lengths and fiber lengths, with the symbol rate set to 32~GBd. The modeled width is based on Equation~(\ref{eq:SpectralDipWidth}).}
  \label{fig:VaryingBL}
\end{figure}

\begin{figure}[!htbp]
  \centering
  \includegraphics[width=0.76\linewidth]{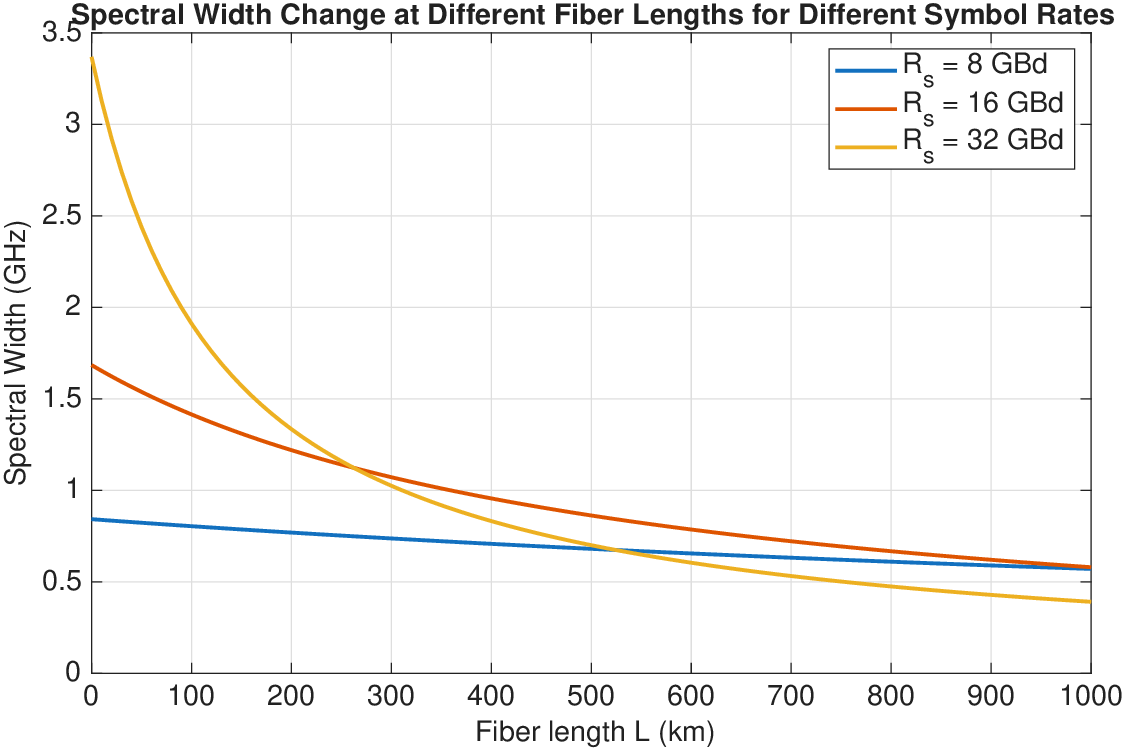}
  \caption{Modeled width of the spectral dip at $f=0$~Hz for different symbol rates, with the block length set to 18 symbols. The modeled width is based on Equation~(\ref{eq:SpectralDipWidth}).}
  \label{fig:VaryingSymbolRate}
\end{figure}

\subsubsection{Dependence on Block Length}
In order to understand how the width of the spectral dip changes with different block lengths, Equation~(\ref{eq:SpectralDipWidth}) can be used. Keeping the symbol rate constant at 32~GBd, the width of the spectral dip is modeled for different block lengths at different fiber lengths. The results are shown in Fig.~\ref{fig:VaryingBL}, where it can be seen that the spectral dip is wider for shorter block lengths than for longer block lengths. However, as the fiber length increases, the spectral dip becomes narrower, and at large fiber distances, the spectral-dip widths for all block lengths converge to a similar value. This is because the dispersion effect causes the pulse to spread out, which reduces the bandwidth of the energy-signal PSD of the envelope.

\subsubsection{Dependence on Symbol Rate}
The width of the spectral dip can also be controlled by adjusting the symbol rate, as described by Equation~(\ref{eq:SpectralDipWidth}). The results are shown in Fig.~\ref{fig:VaryingSymbolRate}, where it can be seen that as the symbol rate increases, the width of the spectral dip becomes wider. However, dispersion causes significant overlap between the pulses at high symbol rates, proportionally stretching the block duration, which can lead to increased low-frequency energy fluctuations. Since dispersion affects different symbol rates differently, it leads to an optimum symbol rate that has the lowest low-frequency fluctuations. Here, the block length has been kept constant at 18 symbols per block while the symbol rate is varied. The results demonstrated here are valid for any block-based shaping method as long as the statistical properties outlined in Equations~(\ref{eq:1DPSD}) and (\ref{eq:2DPSD}) are known.

\subsection{Optimum Symbol Rate}
The optimum symbol rate can be determined from the width of the spectral dip in Equation~(\ref{eq:SpectralDipWidth}). Since the width of the spectral dip is a function of the symbol rate, simple calculus can be used to determine the optimum symbol rate that maximizes the width of the spectral dip. The optimum symbol rate is determined to be
\begin{equation}\label{eq:OptimalSymbolRateShaped}
R_{s,shaped}^{\mathrm{opt}} = \sqrt{\frac{(n_s - 1 + 2\,a)\,c}{{\kappa_\beta}\,D\,L\,\lambda^{2}}} = \sqrt{\frac{n_s - 1 + 2\,a}{2\pi\,L\,|\beta_2|\,{\kappa_\beta}}}
\end{equation}
Here, $\vert\beta_2\vert$ is the group velocity dispersion (GVD) parameter, related to the dispersion coefficient $D$ by $\vert\beta_2\vert = D\lambda^2/(2\pi c)$. 

In the specific case where the block length $n_s = 1$, the model in Eq.~(\ref{eq:BlockDurationAfterDispersion}) reduces to the duration of a single symbol that experiences dispersion. The corresponding energy signal, found via the square law, represents the self-beating of the pulse, which corresponds to the second term in the PSD models of Eqs.~(\ref{eq:1DPSD}) and (\ref{eq:2DPSD}). This special case gives a practical baseline for choosing the symbol rate in an unshaped system. As accumulated dispersion increases, the inter-symbol beat frequency produced by the square law of the energy signal decreases, while the self-beating bandwidth scales with the same dispersion-induced spreading. The optimal symbol rate is therefore the one that maximizes the beat frequency (equivalently, minimizes the effective block duration), which simultaneously widens the self-beating band and suppresses low-frequency energy fluctuations. The optimum symbol rate becomes

\begin{equation}\label{eq:OptimalSymbolRateUnshaped}
R_{s,unshaped}^{\mathrm{opt}} = \sqrt{\frac{2\,a\,c}{{\kappa_\beta}\,D\,L\,\lambda^{2}}} = \sqrt{\frac{a}{\pi\,L\,|\beta_2|\,{\kappa_\beta}}}
\end{equation}

However, two other optimal symbol-rate equations with a form similar to that of the equation derived in (\ref{eq:OptimalSymbolRateUnshaped}) have been proposed in \cite{poggiolini2016analytical} and \cite{wang2017detailed}. In a more generic form, the optimal symbol-rate equation is represented as:

\begin{equation}\label{eq:OptimalGeneral}
    {R_{\text{g}}} = \sqrt{\frac{OL}{2\pi|\beta_2|{L_{\text{Link}}}\nu}}
\end{equation}
where $OL$ is the overlap factor, which is the number of symbols into which a particular symbol of interest spreads within a signal, and $\nu$ is the factor for handling subcarrier spacing. The optimal symbol-rate equation derived in (\ref{eq:OptimalSymbolRateUnshaped}) has $OL = 2\,a$ and $\nu = {\kappa_\beta}$. Poggolini~\emph{et~al.} proposed a similar equation with $OL = 4$ and $\nu = 2\frac{\delta f}{R} - 1$, while Wang~\emph{et~al.} \cite{wang2017detailed} set $OL = 1$ and $\nu = \left(\frac{\delta f}{R}\right)^2$.  

\subsection{Intensity Fluctuations Spectrum}
\begin{figure}[!htbp]
  \centering
  \includegraphics[width=0.76\linewidth]{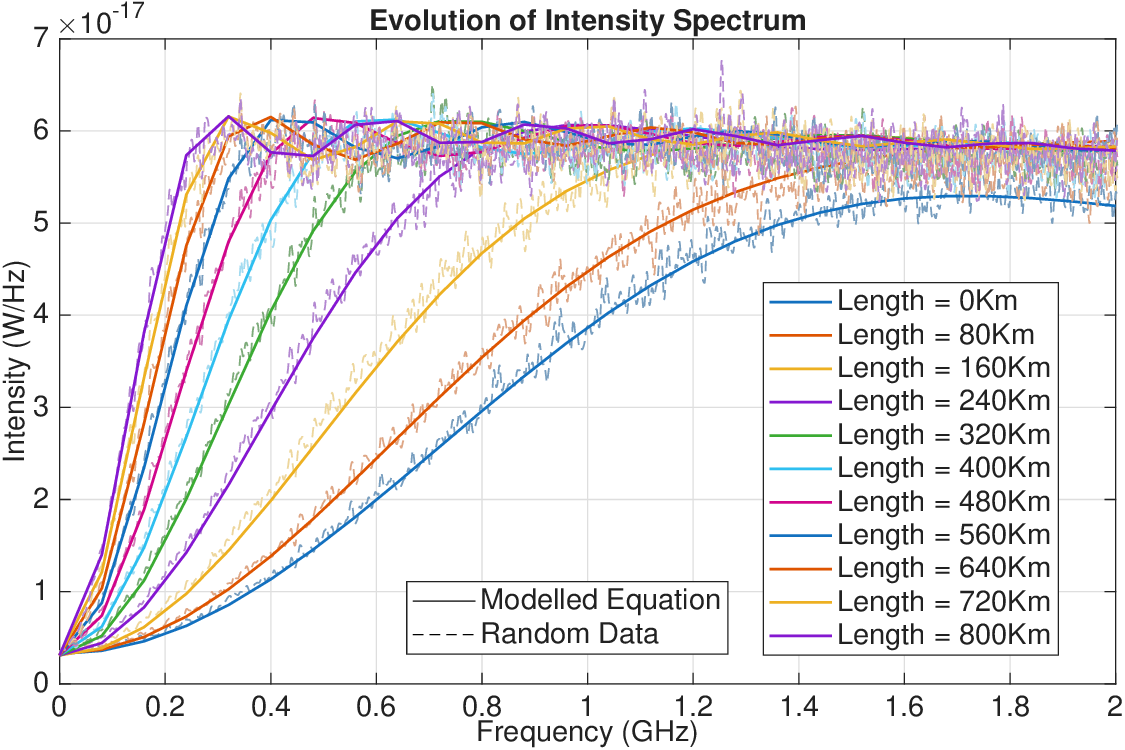}
  \caption{Intensity-fluctuation spectra for a 64-QAM signal transmitted at 32~GBd and shaped by the ESS method with a rate of 4.8 bits/symbol and $n_s = 18$. The modeled intensity-fluctuation spectra are based on the semi-analytical model presented in Equation~(\ref{eq:IntensityFluctuation}) and the results are compared against the VPItransmissionMaker Optical Systems v11.5 simulation results.}
  \label{fig:IntensitySpectra}
\end{figure}

In this section, a full optical simulation is conducted to validate the semi-analytical model presented in Equation~(\ref{eq:IntensityFluctuation}) for the intensity-fluctuation spectrum, which is later used in the XPM phase noise model. Although little difference is expected between the energy-signal PSD and the intensity-fluctuation spectrum, it is necessary to validate the semi-analytical model presented in Equation~(\ref{eq:IntensityFluctuation}), which is a key component of the XPM phase noise model that establishes the relationship between the intensity-fluctuation spectrum and the XPM phase noise. Fig.~\ref{fig:IntensitySpectra} presents the modeled spectra of intensity fluctuations at different fiber lengths. The results are for a 64-QAM signal transmitted at 32~GBd and shaped by the ESS method with a rate of 4.8 bits/symbol and $n_s = 18$. Unlike the CCDM method, the ESS algorithm provides additional control of the spectral dip through the shaping block length, as shown in Fig.~\ref{fig:psd-comparison}, and the spectral dip at $f = 0$ Hz is not suppressed. The modeled intensity-fluctuation spectra are based on the semi-analytical model presented in Equation~(\ref{eq:IntensityFluctuation}) and are compared against the simulation results. The modeled intensity-fluctuation spectra show good agreement with the VPItransmissionMaker Optical Systems v11.5 simulation results, which validates the semi-analytical model presented in Equation~(\ref{eq:IntensityFluctuation}). As suggested in \cite{Prasad2025intenstiy}, the intensity-fluctuation spectra at low frequencies are dominated by the beat between neighboring symbols, which leads to an increase in intensity fluctuations at low frequencies, and shaped constellations are no exception. 

\subsection{XPM Phase Noise}
\subsubsection{Effect of Shaping Methods}

\begin{figure}[!htbp]
  \centering
  \begin{subfigure}[t]{0.76\linewidth}
    \centering
    \includegraphics[width=\linewidth]{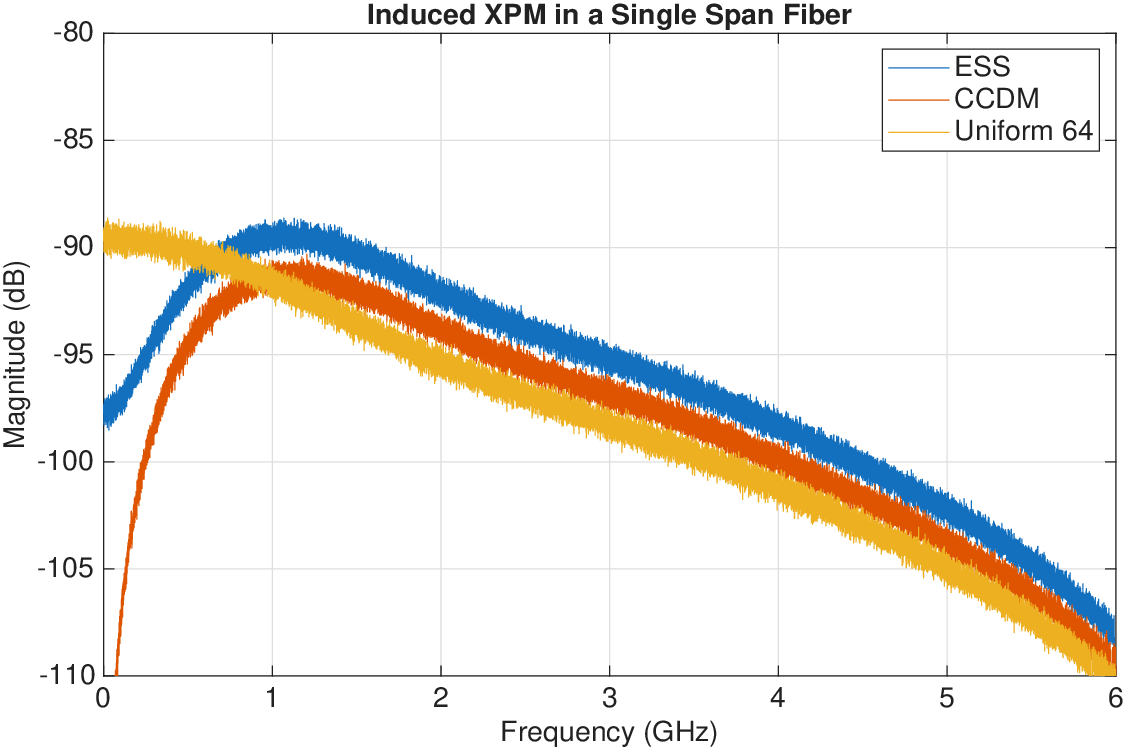}
  \end{subfigure}
  \hfill
  \begin{subfigure}[t]{0.76\linewidth}
    \centering
    \includegraphics[width=\linewidth]{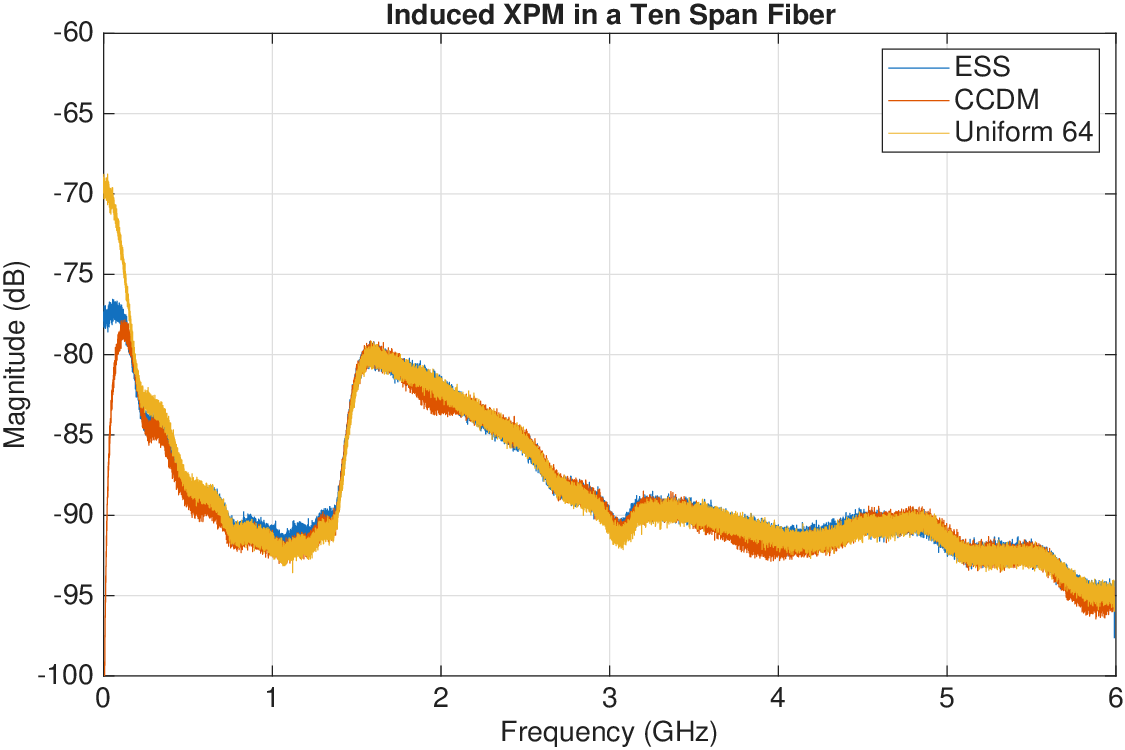}
  \end{subfigure}
  \caption{Phase noise induced on a 10 $\mu$W probe channel located 50 GHz from a pump channel with a channel power of 1 mW. The phase noise is shown for single-span (top) and 10-span (bottom) fiber systems, demonstrating the filter-like behavior of the XPM transfer function.}
  \label{fig:PNSpectra}
\end{figure}

In the XPM model \cite{lowery2022XPM,chiang1994cross1,chiang1996cross}, the phase-noise power spectral density (PSD) is driven by the PSD of data-induced intensity fluctuations. In particular, growth at low frequencies (Fig.~\ref{fig:IntensitySpectra}) degrades system performance. Accounting for this growth enables the model to reproduce the measured phase-noise spectra and to explicitly relate the intensity-fluctuation spectrum from one span to the next and the resulting XPM phase noise. Consistent with this, Fig.~\ref{fig:PNSpectra} compares phase-noise spectra for different shaping methods and shows that the low-frequency region is strongly method-dependent. For a single-span fiber, the wideband nature of the XPM efficiency function \cite{chiang1996cross} makes the entire bandwidth relevant. However, as the number of spans increases, the contribution of low frequencies becomes dominant \cite{amari2019introducing,peng2021baud}. In the ten-span system phase-fluctuation spectrum shown in Fig.~\ref{fig:PNSpectra}, two lobes appear: a sinc-like envelope arises from the coherent vector summation of span-by-span intensity fluctuations and leads to the link efficiency factor \cite{lowery2022XPM,chiang1996cross}; when this envelope is combined with the XPM-efficiency curve, the first lobe dominates the phase-noise power. The broader second lobe reflects the finite pass-band of the pump channel, where contributions from a range of frequency separations accumulate and produce a wider response \cite{lowery2022XPM}.

\begin{figure}[!htbp]
  \centering
  \includegraphics[width=0.76\linewidth]{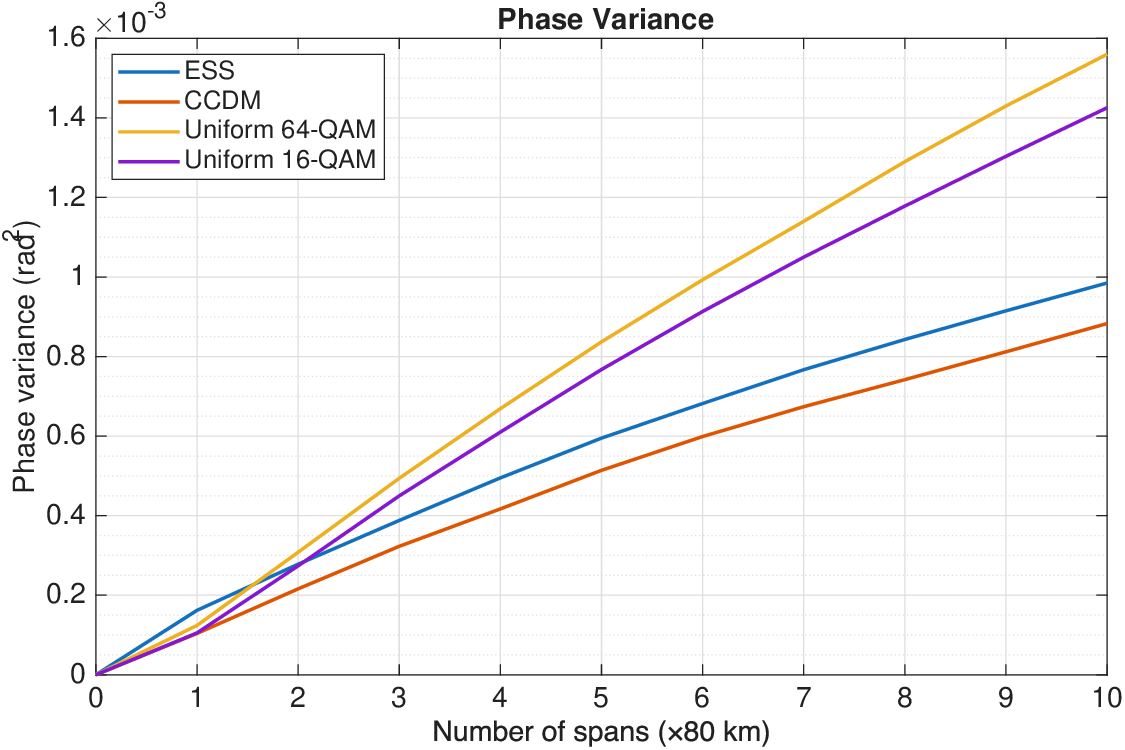}
  \caption{Comparison of phase variance for unshaped 16-QAM, and 64-QAM signals together with ESS- and CCDM-shaped 64-QAM signals at a shaping rate of 4.8~bits/symbol and a block length of 18 symbols.}
  \label{fig:PNShapingMethod}
\end{figure}

Figure~\ref{fig:PNShapingMethod} compares the phase variance measured on a probe channel for unshaped 64-QAM, unshaped 16-QAM, ESS-shaped 64-QAM, and CCDM-shaped 64-QAM. The lower phase variance obtained with CCDM is consistent with its stronger suppression of the near-DC IF pedestal. ESS provides substantial low-frequency attenuation but retains residual near-DC energy at moderate $n_s$, which explains its slightly higher phase variance relative to CCDM.

For the phase-variance comparison in Fig.~\ref{fig:PNShapingMethod}, all pump-channel waveforms are evaluated at the same launch power of $1$~mW, the same 32~GBd symbol rate, the same 50~GHz channel spacing, and the same fiber-link parameters. The shaped ESS and CCDM cases use 64-QAM at 4.8~bits/symbol with $n_s=18$. The unshaped 64-QAM and 16-QAM are not intended to be equal-information-rate references; rather, they provide equal-launch-power baselines with different symbol-energy variances. The comparison therefore isolates how the energy statistics and low-frequency IF spectra affect XPM-induced phase variance.

ESS exhibits a slightly higher phase variance than CCDM, attributable to residual low-frequency fluctuations in the intensity-fluctuation spectrum. With a finite block length $n_s$, ESS permits inter-block variability in the mean energy, so the variance of the block mean is nonzero ($\sigma_{\mu_{\mathrm{blk}}}^{2}>0$); this leaves a finite DC component that CCDM’s constant composition suppresses by construction ($\sigma_{\mu_{\mathrm{blk}}}^{2}=0$). Although ESS attenuates low-frequency content, it does not fully null the DC component for moderate $n_s$, particularly when $\mu_{\sigma_{\mathrm{blk}}^{2}}\neq\sigma_{E}^{2}$ with $\sigma_{E}^{2}>0$. As $n_s$ increases, with the energy threshold re-tuned to maintain the shaping rate, the block-energy distribution concentrates ($\sigma_{\mu_{\mathrm{blk}}}^{2}\to 0$) and the low-frequency pedestal diminishes; in the large-$n_s$ regime the phase-variance gap between ESS and CCDM would become negligible. In practice, the choice of $n_s$ trades residual low-frequency energy, and thus phase variance, against rate loss and computational complexity, which explains the small but consistent excess of ESS over CCDM in Fig.~\ref{fig:PNShapingMethod}.

Over the first few spans, ESS provides little or no reduction in probe channel phase variance relative to unshaped 64-QAM and 16-QAM. At short distances, shaping increases the symbol energy variance, which raises the intensity fluctuation spectrum above the unshaped baselines, in line with Equations~(\ref{eq:1DPSD}) and (\ref{eq:2DPSD}). As the number of spans grows, the balance reverses: the XPM transfer behaves effectively as a low-pass filter \cite{askari2024probabilistic,peng2021baud}, so low-frequency components dominate the induced phase noise. By attenuating these components, especially near DC, ESS drives the phase variance below the unshaped curves after approximately two spans (Fig.~\ref{fig:PNShapingMethod}). The degree of low-frequency attenuation depends on $n_s$, $\mu_{\sigma_{\mathrm{blk}}^{2}}$, and $\sigma_{\mu_{\mathrm{blk}}}^{2}$; with appropriate parameter choices, further performance gains are achievable. Finally, because symbol energy variance increases with constellation order, the unshaped 64-QAM reference exhibits a larger intensity fluctuation spectrum and therefore higher phase variance than the unshaped 16-QAM reference.

\begin{figure}[!htbp]
  \centering
  \begin{subfigure}[t]{0.76\linewidth}
    \centering
    \includegraphics[width=\linewidth]{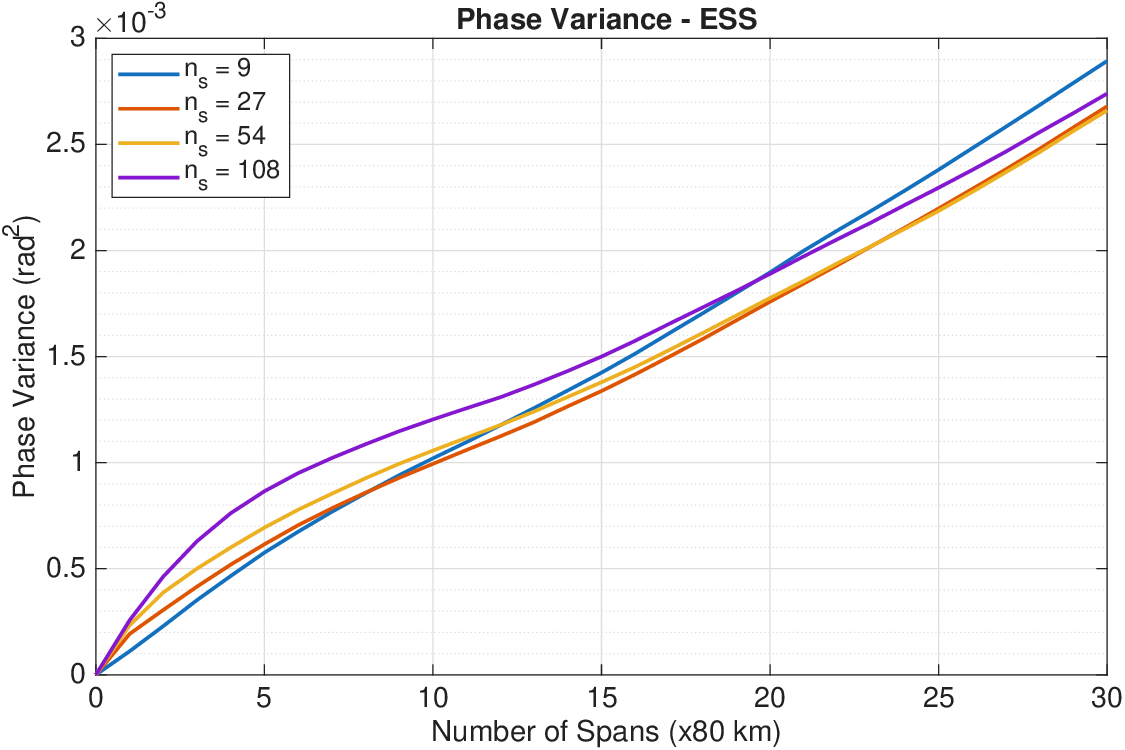}
  \end{subfigure}
  \hfill
  \begin{subfigure}[t]{0.76\linewidth}
    \centering
    \includegraphics[width=\linewidth]{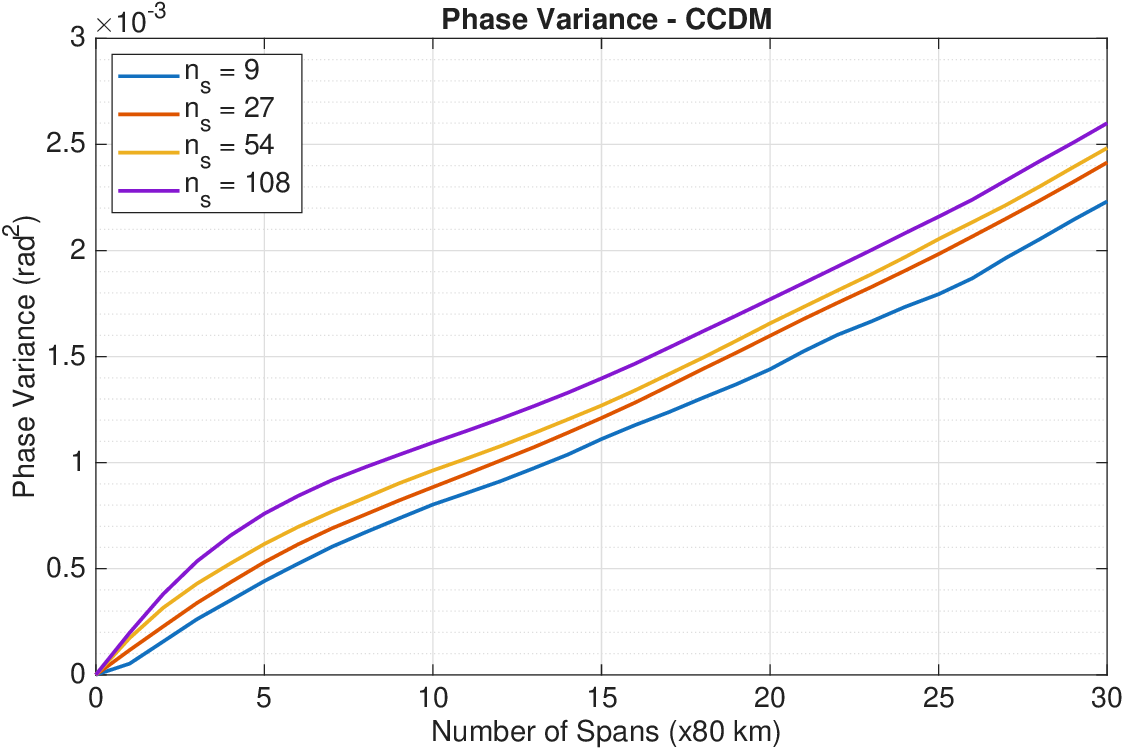}
  \end{subfigure}
  \caption{Phase-noise variance induced on a probe channel by ESS- and CCDM-shaped 64-QAM pump signals for different block lengths at a shaping rate of 4.8~bits/symbol.}
  \label{fig:BLPNComparison}
\end{figure}

\subsubsection{Effect of Block Length}
As established by Equation~(\ref{eq:SpectralDipWidth}) and Fig.~\ref{fig:VaryingBL}, increasing the block length narrows the spectral dip, whereas decreasing the block length widens it. Figure~\ref{fig:BLPNComparison} shows the phase noise induced on the probe channel for different block lengths using ESS- and CCDM-shaped 64-QAM at a shaping rate of 4.8~bits/symbol. Increasing $n_s$ initially increases the phase-noise variance over the first several spans, before an optimum block length emerges at longer distances. For ESS, shorter blocks widen the spectral dip but can also reduce its depth, leaving greater residual near-DC intensity fluctuations. In comparison with CCDM, the spectral dip mainly changes in width while the near-DC suppression remains stronger because the constant-composition constraint fixes the block composition. Therefore, the phase variance depends not only on the spectral-dip width but also on the near-DC dip depth, which explains the optimum block length observed for ESS-shaped signals.

\subsubsection{Effect of Symbol Rate}
\begin{figure}[!htbp]
  \centering
  \includegraphics[width=0.76\linewidth]{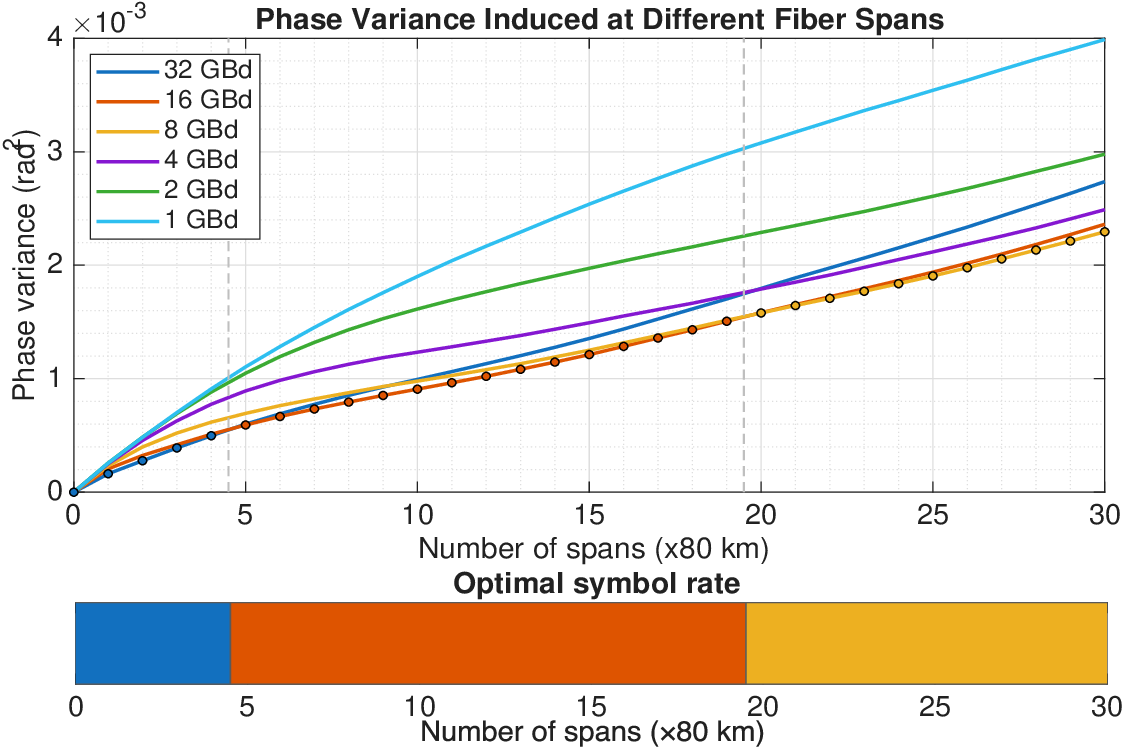}
  \caption{Phase-noise variance induced on a probe channel by an ESS-shaped 64-QAM pump for different symbol rates, at a shaping rate of 4.8~bits/symbol and $n_s=18$.}
  \label{fig:OptimumESS}
\end{figure}

\begin{figure}[!htbp]
  \centering
  \includegraphics[width=0.76\linewidth]{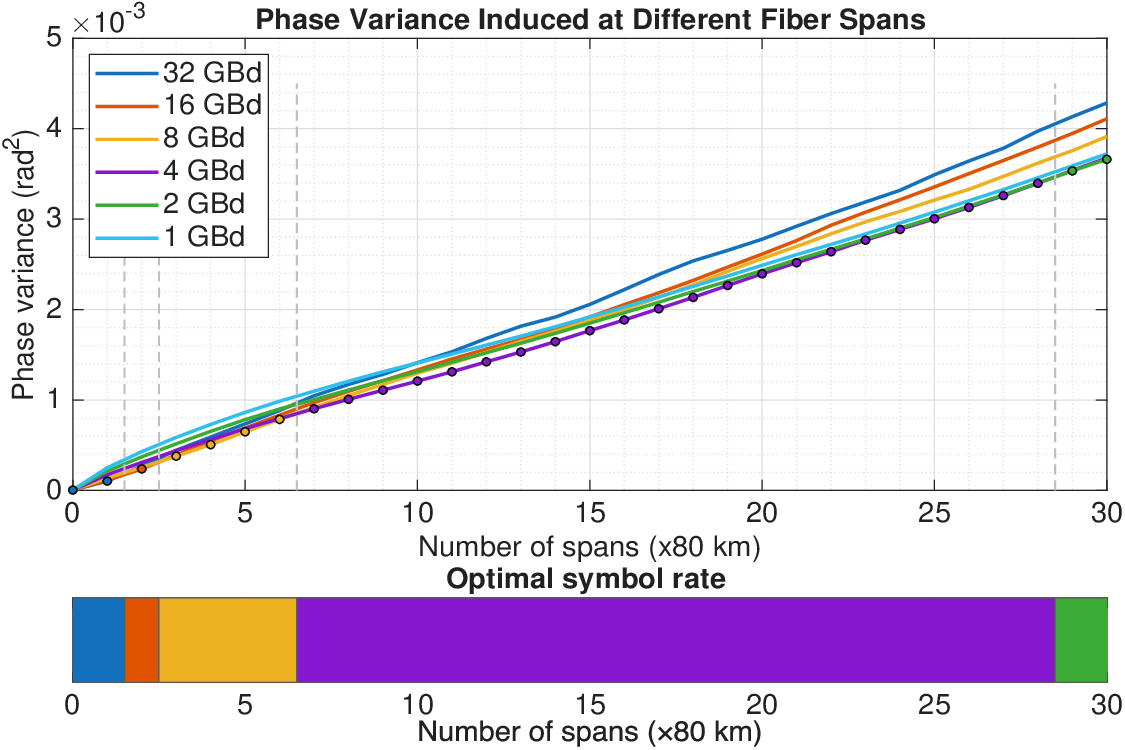}
  \caption{Phase-noise variance induced on a probe channel by an unshaped 16-QAM pump for different symbol rates.}
  \label{fig:OptimumUnshaped}
\end{figure}

The symbol rate significantly affects the spectral dip near $f=0$~Hz and, consequently, the XPM-induced phase noise variance~\cite{Prasad2025intenstiy}. At shorter transmission distances, a lower symbol rate produces stronger low-frequency intensity fluctuations, leading to higher phase noise on a co-propagating probe channel. However, as a signal propagates, chromatic dispersion broadens the pulses, causing an additional build-up of low-frequency spectral components, an effect known as chirp-induced intensity fluctuations~\cite{Prasad2025intenstiy} (Figs.~\ref{fig:psddistance-comparison} and \ref{fig:IntensitySpectra}). The interplay between these two effects---intrinsic intensity fluctuations (dominant at low $R_s$ and short distances) and chirp-induced fluctuations (dominant at high $R_s$ and long distances)---gives rise to an optimum symbol rate that minimizes the total low-frequency noise power \cite{lowery2022XPM,du2011optimizing}. Since XPM-induced phase noise is directly proportional to these intensity fluctuations, this optimum symbol rate also induces the least phase noise variance on the probe channel. This trend is demonstrated in Figs.~\ref{fig:OptimumESS} and \ref{fig:OptimumUnshaped}, which show the phase noise variance versus symbol rate for different transmission lengths. Fig.~\ref{fig:OptimumESS} corresponds to a pump channel shaped using the ESS method (4.8~bits/symbol, $n_s=18$), while Fig.~\ref{fig:OptimumUnshaped} corresponds to an unshaped 16-QAM pump. Both scenarios exhibit the same behavior: as the transmission length increases, the overall phase noise variance grows, and the optimal symbol rate (which induces the minimum phase noise) shifts to a lower value. 

\begin{figure}[!htbp]
  \centering
  \includegraphics[width=0.76\linewidth]{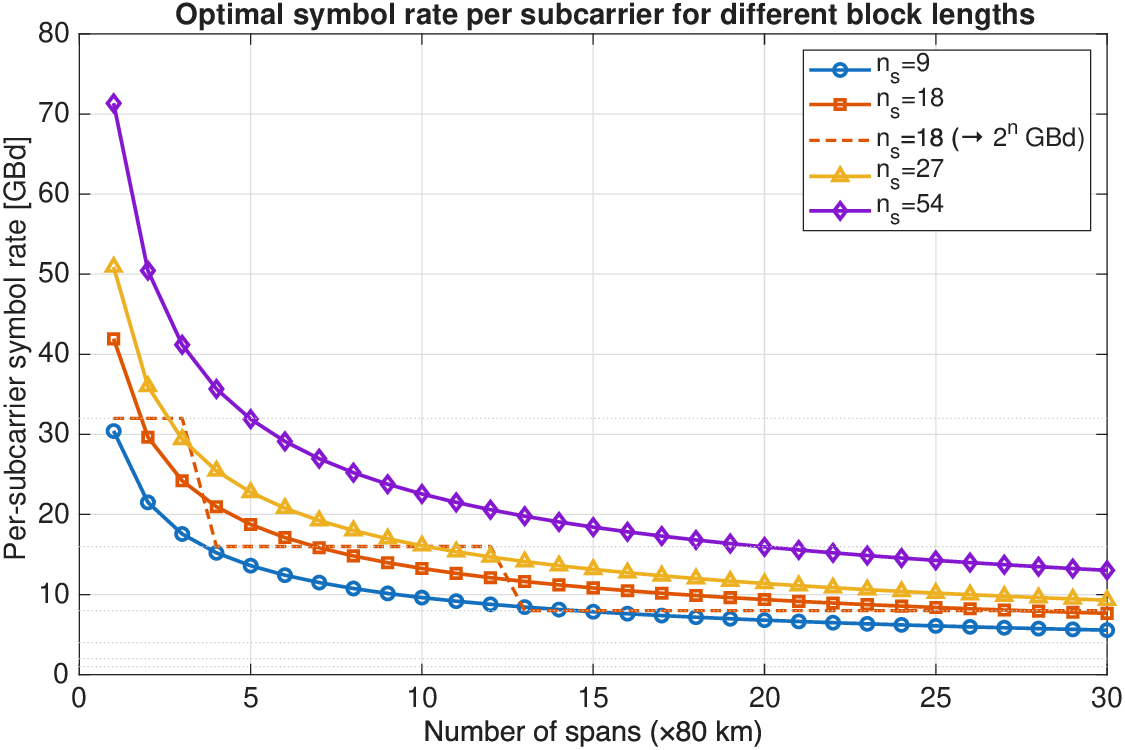}
  \caption{Optimal symbol rate obtained from Equation~(\ref{eq:OptimalSymbolRateShaped}) with $a\approx1$ for different block lengths.}
  \label{fig:ModelOptimumShaped}
\end{figure}

Furthermore, Figs.~\ref{fig:OptimumESS} and \ref{fig:OptimumUnshaped} show that introducing block-based shaping increases the optimum symbol rate. This means that shaping allows data transmission at higher symbol rates over longer transmission distances while maintaining a lower phase noise variance, as shown in Figs.~\ref{fig:OptimumESS} and \ref{fig:OptimumUnshaped}. For $n_s = 18$, the optimum symbol rate determined by minimizing the phase variance in Fig.~\ref{fig:OptimumESS} agrees well with the modeled optimum symbol rate derived in Equation~(\ref{eq:OptimalSymbolRateShaped}) and shown in Fig.~\ref{fig:ModelOptimumShaped}. For example, for $N$ = 3, 10, and 27, Fig.~\ref{fig:ModelOptimumShaped} gives corresponding optimum symbol rates of approximately 32~GBd, 16~GBd, and 8~GBd, respectively, which correspond to the optimum symbol rates identified in Fig.~\ref{fig:OptimumESS} for $n_s$ = 18. This indicates that the modeled optimum symbol rate is valid for shaped signals. 

Also, based on Fig.~\ref{fig:ModelOptimumShaped}, it is observed that the optimum symbol rate increases with block length. This behavior is expected based on Equations (\ref{eq:BlockDurationAfterDispersion}) and (\ref{eq:SpectralDipWidth}); longer blocks enlarge the intrinsic low-frequency intensity fluctuations of the energy signal, whereas higher symbol rates generate stronger chirp-induced fluctuations through chromatic-dispersion broadening. The operating point that minimizes phase-noise impact is the balance between these two mechanisms; increasing $n_s$ therefore pushes the optimum $R_s$ upward and keeps that higher rate optimal over a longer distance. As distance grows, the optimum rate for every $n_s$ gradually decreases, and the gap between block lengths narrows and converges to the same optimum, indicating a diminishing benefit from selecting a particular $n_s$ in the long-haul regime. In practical systems constrained by a maximum transmission bandwidth, the choice of block length becomes a critical factor. A larger block may not operate at its optimal point because its required optimal symbol rate could demand a total signal bandwidth that exceeds the limitations of the system's hardware, forcing the use of a suboptimal, lower symbol rate.

\begin{figure}[!htbp]
  \centering
  \includegraphics[width=0.76\linewidth]{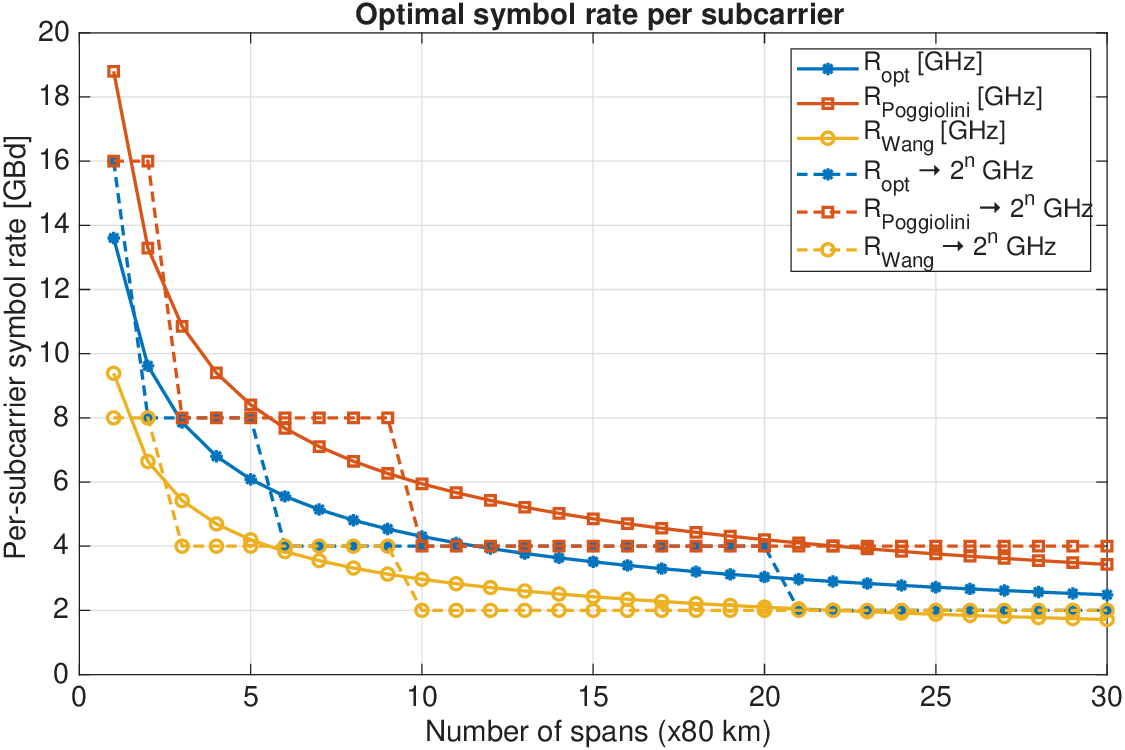}
  \caption{Optimal symbol rate obtained from Equation~(\ref{eq:OptimalSymbolRateUnshaped}) with $a\approx1$ and compared with the optimal-symbol-rate expressions derived by Poggiolini~\emph{et~al.} \cite{poggiolini2016analytical}, $R_{\mathrm{Poggiolini}}$, and Wang~\emph{et~al.} \cite{wang2017detailed}, $R_{\mathrm{Wang}}$.}
  \label{fig:ModelOptimumUnshaped}
\end{figure}

For unshaped signals, the optimum symbol rate determined through Equation~(\ref{eq:OptimalSymbolRateUnshaped}) follows the optimum symbol-rate trend demonstrated in Fig.~\ref{fig:OptimumUnshaped} more closely than the optimum equations presented in \cite{poggiolini2016analytical} and \cite{wang2017detailed}. For example, in Fig.~\ref{fig:ModelOptimumUnshaped}, $N$ = 3, 11, and 29 correspond to approximate optimum symbol rates of 8~GBd, 4~GBd, and 2~GBd, respectively, in good agreement with the optimum symbol rates identified in Fig.~\ref{fig:OptimumUnshaped}. Based on Fig.~\ref{fig:ModelOptimumUnshaped}, it is observed that the presented equations are a middle ground between the optimum symbol-rate equations presented in \cite{poggiolini2016analytical} and \cite{wang2017detailed}. It should be emphasized that the phase-variance curves used to select the optimum in Figs.~\ref{fig:OptimumESS} and \ref{fig:OptimumUnshaped} are tightly grouped; near the transition regions adjacent rates produce nearly identical variance. Small numerical or measurement uncertainties (finite Monte Carlo averaging, PSD resolution, and discrete-rate rounding) can therefore flip the selected optimum between neighboring grid values. For design purposes, the optimum symbol rate is best interpreted as a narrow range around the theoretical prediction rather than a single exact value.

\FloatBarrier
\section{Conclusion}
This work establishes intensity-fluctuation spectra as a practical design tool for constellation shaping under fiber nonlinearity. Starting from an analytical decomposition of the energy-signal PSD, explicit terms were identified that govern low-frequency content: the self-beating contribution, inter-symbol beating across symbol lags, and the block-structure term that depends on the average within-block energy variance $\mu_{\sigma_{\mathrm{blk}}^{2}}$ and the variance of per-block mean energy $\sigma_{\mu_{\mathrm{blk}}}^{2}$. These terms explain the contrasting behaviours of CCDM and ESS at $f = 0$ Hz: CCDM’s constant composition suppresses the DC pedestal, whereas ESS exhibits a finite pedestal for moderate block lengths, diminishing as block energy concentrates with increasing $n_s$.

A simple width law for the DC spectral dip, $\Delta f_b = 2/T_b'$, was derived, where $T_b'$ combines block duration, pulse roll-off, and dispersion. This relation guides co-optimization of block length and symbol rate: shorter blocks and moderate rates widen the dip at short distances, while dispersion narrows it with propagation. From this trade-off, closed-form optimal symbol-rate expressions were obtained for shaped and unshaped systems, matching the minima of XPM-induced phase variance observed in simulations. The analysis predicts (and simulations confirm) that shaping lowers phase variance beyond a few spans, with CCDM offering the strongest low-frequency suppression and ESS approaching it as $n_s$ grows; an ESS optimum block length emerges from the joint control of dip width and depth.

These results yield concrete rules for shaping based on spectral composition to minimise the low-frequency intensity fluctuations that dominate XPM, unifying time-domain energy statistics with frequency-domain PSD features. An explicit symbol-rate rule is provided to achieve the lowest XPM phase noise for both shaped and unshaped systems. The framework readily extends to other shaping methods and modulation formats, offering a practical design pathway for nonlinear-tolerant optical transmission. Future work includes extending the framework to multi-channel interactions beyond XPM (e.g., SPM/FWM coupling), polarization-multiplexed 4D shaping with explicit covariance control, integration with GN/EGN-style system models including ASE noise, and realization of adaptive, distance-aware shaping that tracks dispersion accumulation in real time.

\section*{Acknowledgment}
The authors wish to express sincere gratitude to Professor Arthur Lowery for his invaluable guidance, insightful discussions, and constructive feedback throughout the course of this research. This work was supported by the Faculty of Engineering Publication Award, Monash University.

\bibliographystyle{elsarticle-num}
\bibliography{Ref}

\end{document}